\documentclass[preprint,12pt,authoryear]{elsarticle}

\usepackage{subfiles}
\usepackage{amssymb}
\usepackage{amsmath}
\usepackage{subfig}
\usepackage{booktabs}
\usepackage{mathtools}
\usepackage{makecell}
\usepackage{color, soul}
\usepackage{siunitx}
\usepackage{diffcoeff}
\usepackage[final]{changes}

\setdeletedmarkup{\color{red}\sout{#1}}
\setcitestyle{square,comma,numbers,sort&compress}
\newcommand{\dint}{\,\mathrm{d}}

\journal{High Energy Density Physics}

\begin{document}
\definechangesauthor[name=1]{P1}
\definechangesauthor[name=2]{P2}
\definechangesauthor[name=3]{P3}
\definechangesauthor[name=4]{P4}
\definechangesauthor[name=5]{P5}
\definechangesauthor[name=6]{P6}
\definechangesauthor[name=7]{P7}
\definechangesauthor[name=8]{P8}
\definechangesauthor[name=9]{P9}
\definechangesauthor[name=10]{P10}
\definechangesauthor[name=11]{P11}

\begin{frontmatter}

\title{FLAIM: a reduced volume ignition model for the compression and thermonuclear burn of spherical fuel capsules}

\author[a]{Abd Essamade Saufi\corref{cor}}
\cortext[cor]{Corresponding author}
\ead{abdessamade.saufi@firstlightfusion.com}
\author[a]{Hannah Bellenbaum \fnref{1}}
\fntext[1]{Present address: Center for Advanced Systems Understanding (CASUS), D-02826 Görlitz, Germany; Helmholtz-Zentrum Dresden-Rossendorf (HZDR), D-01328 Dresden, Germany}
\author[a]{Martin Read}
\author[a]{Nicolas Niasse}
\author[a]{Sean Barrett}
\author[a]{Nicholas Hawker}
\author[a]{Nathan Joiner}
\author[a]{David Chapman}
\affiliation[a]{organization={First Light Fusion Ltd.},
             addressline={Unit 9/10, Oxford Pioneer Park, Mead Road},
             city={Oxford},
             postcode={OX5 1QU},
             state={Oxfordshire},
             country={United Kingdom (UK)}}

\begin{abstract}
We present the ``First Light Advanced Ignition Model" (FLAIM), a reduced model for the implosion, adiabatic compression, volume ignition and thermonuclear burn of a spherical DT  fuel capsule utilising a high-Z metal pusher. \deleted[id=P3]{FLAIM adopts state-of-the-art solutions for the equation of state, transport properties, wall model and burn scheme.} \replaced[id=P3]{\texttt{FLAIM}}{It} is characterised by a highly modular structure, which makes it an appropriate tool for optimisations, sensitivity analyses and parameter scans. One of the key features of the code is the 1D description of the hydrodynamic operator, which has a minor impact on the computational efficiency, but allows us to gain a major advantage in terms of physical accuracy. \replaced[id=P1]{We demonstrate that a more accurate treatment of the hydrodynamics plays a primary role in closing most of the gap between a simple model and a general 1D rad-hydro code, and that only a residual part of the discrepancy is attributable to the heat losses.}{We demonstrate the primary role that a correct treatment of the hydrodynamics plays in closing most of the gap between a simple model and a general high-fidelity code, and how the detailed modelling of other physical phenomena (e.g.,  heat losses) is of secondary importance.} We present a detailed quantitative comparison between FLAIM and \replaced[id=P2]{1D rad-hydro}{full-physics 1D hydrodynamic} simulations, showing \replaced[id=P2]{good}{excellent} agreement over a large parameter space in terms of temporal profiles of key physical quantities, ignition maps and typical burn metrics.
\end{abstract}

\end{frontmatter}


\section{Introduction}
\label{introduction}
\noindent First Light Fusion (FLF) is a privately-funded research company working on alternative solutions for inertial confinement fusion (ICF) \cite{flf_website}. \deleted[id=P4]{Our approach is based on a one-sided impact between  a  hyper-velocity projectile and an advanced target, composed of an amplifier and a fuel capsule. The amplifier is specifically designed to focus the delivered kinetic energy on the fuel capsule, boosting at the same time the impact pressure and the final implosion velocity.} Part of FLF’s current effort is devoted to \replaced[id=P5]{the development of numerical capabilities}{the development of \texttt{FuSE} (Fusion System Evaluator), an  end-to-end fusion systems code employed.} to quickly iterate \replaced[id=P5]{and perform sensitivity analyses on}{and optimise} the design of our drivers \added[id=P5]{and targets, which requires the systematic exploration of} \deleted[id=P5]{and to efficiently explore} a  vast multi-dimensional parameter space at low cost. \deleted[id=P5]{For this reason, computational efficiency is of principal importance.} \replaced[id=P5]{A fundamental step is the development of}{It is fundamental to develop} \textit{simplified} models, that can run rapidly, and whose aim is to capture in broad terms the \replaced[id=P5]{physical phenomena}{main physics} involved \replaced[id=P5]{in a general end-to-end system (e.g., from the driver physics to fuel ignition and burn).}{in each \texttt{FuSE} component, namely:  (i) the launcher, (ii) the projectile dynamics, (iii) the amplifier physics and (iv) the thermonuclear burn of the fusion fuel.} \added[id=P8]{The   objective is to adopt these models as \textit{auxiliary} tools, to help inform the approximate shape of the hyper-surface of a general performance metric (e.g., neutron yield), immediately identify non-viable configurations and eventually  refine the analysis with standard rad-hydro simulations focusing on the most relevant regions.}

The platform is based on \textit{volume ignition}, in particular on the Revolver concept \cite{Molvig_PhysRevLett_2016, Molvig_PhysPlasmas_2018, Keenan_PhysPlasmas_2020b}: using a high-Z metal pusher to confine the fuel, instead of the cold DT shell typical of hot-spot ignition designs, one sacrifices high-gain for robustness, i.e., lower losses, lower ignition temperature and lower implosion velocity \cite{lackner1994equilibrium}. Whilst simple reduced models for hot-spot ignition are numerous, analogous models for volume ignition are less common \cite{Kirkpatrick_NuclFusion_1979, Kirkpatrick_NuclFusion_1981b, Huang_PhysPlasmas_2017, Dodd_PhysPlasmas_2020}\replaced[id=P6]{, }{. Moreover, these models are all inevitably characterised by some form of assumption (e.g., ideal gas behaviour, infinitely thin or constant thickness pusher, thermodynamic equilibrium between the radiation field and the wall)} and frequently lack robust  comparison against high-fidelity \replaced[id=P6]{simulations}{models} in order to assess their accuracy. \deleted[id=P6]{or to support the suitability of such assumptions.}

In this context, the objective of this work is two-fold. Firstly, we present the “First Light Advanced Ignition Model” (\texttt{FLAIM}), a reduced model describing the implosion, volume ignition and thermonuclear burn of a spherical fuel capsule compressed by a high-Z pusher. It includes the use of non-ideal  tabulated equations of state (EoS) and transport properties, 1D treatment of the hydrodynamics, a \replaced[id=P3]{novel approach for the wall modelling}{state-of-the-art model of the wall treatment} \added[id=P6]{, which removes the assumption of thermodynamic equilibrium between the wall and the radiation field, } and a \deleted[id=P7]{complete} thermonuclear burn operator. From an implementation perspective, \texttt{FLAIM}'s main strengths are (i) flexibility, (ii) modularity of the code structure and (iii) computational efficiency, which makes it suitable for large parameter scans and sensitivity analyses. 
Secondly, we present a thorough analysis on the accuracy of \texttt{FLAIM} with respect to high-fidelity 1D simulations in a large design space. In particular, we  highlight the primary role that the hydrodynamic has on the physics of the system, as well as the negative impact that approximate treatments of the hydrodynamics may have on the predictive capability of the model when used to explore ignition regions.

The paper's organisation reflects this purpose, including: a detailed description of the mathematical model and the physical operators  (Section \ref{section:mathematical_model}) and a brief presentation on the numerical methodology, code structure and implementation (Section \ref{section:numerical_setup}). Section \ref{section:comparison_b2} presents a detailed investigation on the accuracy of \texttt{FLAIM}'s hydrodynamic operator by means of cross-code comparison with our in-house hydrodynamic code \texttt{B2}. The analysis is completed in Sections \ref{sec:multi_physics_simulations}-\ref{sec:revolver} by considering the accuracy of \replaced[id=P2]{simulations of igniting targets}{full-physics simulations}, and demonstrating the reliability of \texttt{FLAIM} as a reduced ignition model, comparing the main key diagnostic quantities for the Revolver design with reference results reported in literature.

\section{Mathematical model}
\label{section:mathematical_model}
\noindent In this section, the mathematical model implemented in \texttt{FLAIM} is derived and presented. Figure \ref{pic:main_quantities_and_geometry} represents the main physical and geometric quantities defining the system: the fuel is a DT mixture (with number densities $n_\mathrm{D}, n_\mathrm{T}$) described using a 3-temperature approach (ion $T_\mathrm{i}$, electron $T_\mathrm{e}$, radiation $T_\mathrm{r}$). Based on results from 1D simulations, the high-Z pusher material can be sensibly assumed to have a single temperature $T_\mathrm{p}$. These physical quantities are to be interpreted as \textit{volume-averaged} over the domain of interest, defining a 0D model.  The relevant conserved quantities are evolved over time by physical operators comprising (i) hydrodynamics, (ii) thermal equilibration, (iii) wall, (iv) thermal conduction, (v) radiation loss and (vi) thermonuclear burn. All of the operators have been independently verified against simple benchmark test cases and tested for global mass and energy conservation.

\subsection{Hydrodynamics}
\begin{figure}
	\centering
	\subfloat[]
	{\includegraphics[width=.49\textwidth,height=0.28\textheight]{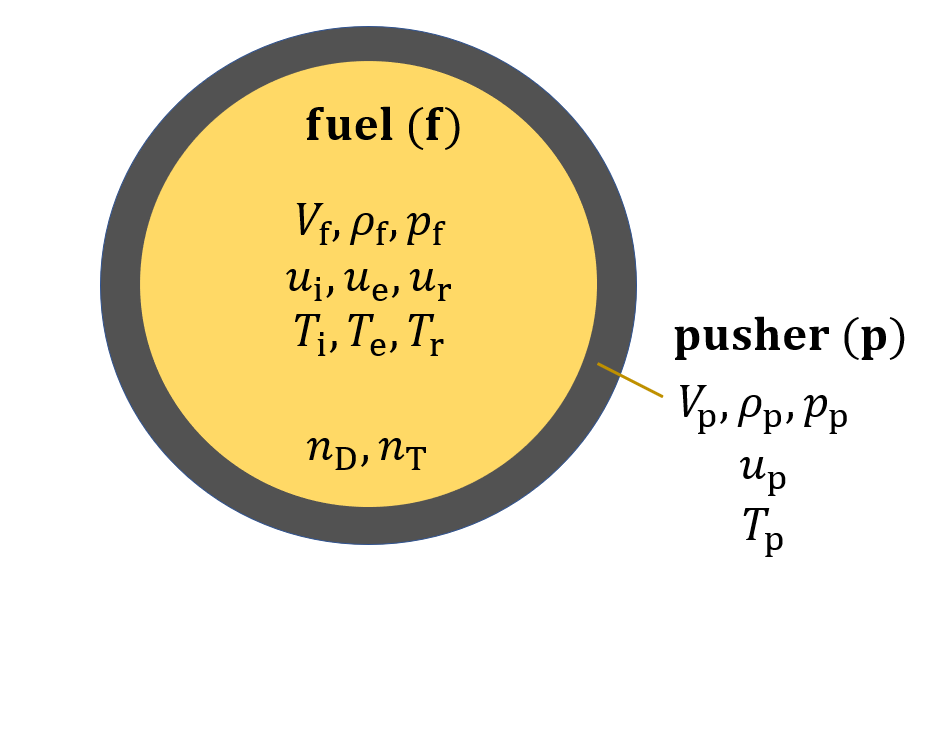}} ~
	\subfloat[]
	{\includegraphics[width=.49\textwidth,height=0.28\textheight]{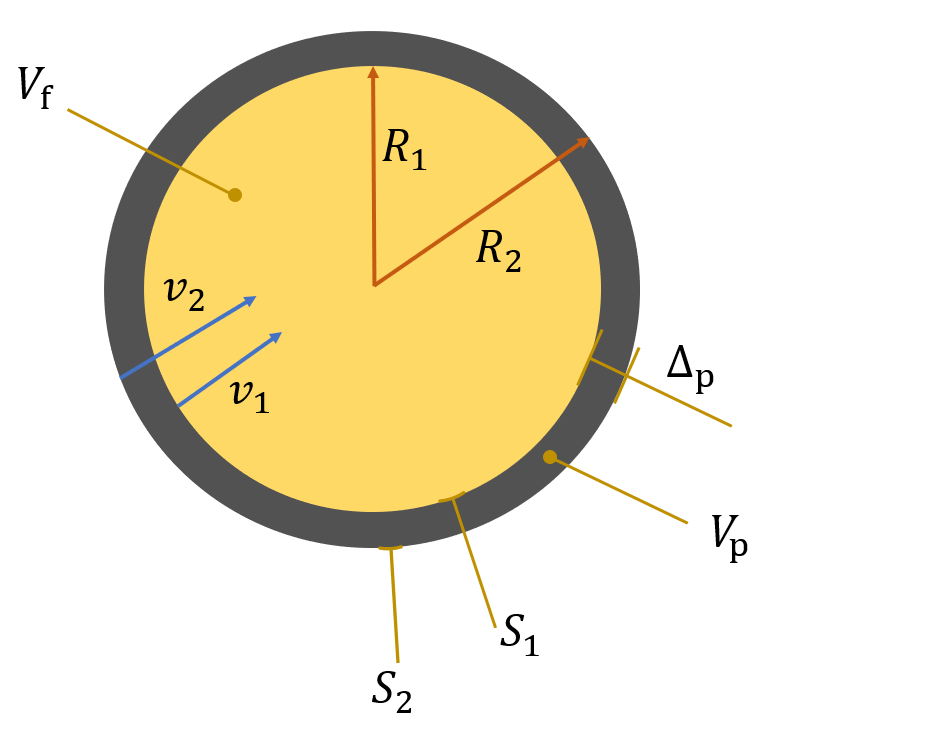}}
	\caption{Main physical and geometrical quantities defining the closed system fuel + pusher.}
	\label{pic:main_quantities_and_geometry}
\end{figure}

\noindent Focusing on the fuel region first (Figure \ref{pic:main_quantities_and_geometry}), we consider a spherical volume $V_\mathrm{f}$ enclosed by a surface $S_1$ on which a pressure $p_f=p_\mathrm{i}+p_\mathrm{e}+p_\mathrm{r}$ is exerted. The following equations describe the temporal variation of total mass, $N_\mathrm{S}$ species number, momentum and internal energy in the system:

\begin{equation}
	\diff{}{t} \int_{V_\mathrm{f}}\rho_\mathrm{f} \dint V_\mathrm{f} = 0 \,,
	\label{continuity_equation}
\end{equation}

\begin{equation}
	\diff{}{t} \int_{V_\mathrm{f}}^{} n_j \dint V_\mathrm{f} = 0 \quad \mathrm{for}~j=0,\dots, N_\mathrm{S}\,,
	\label{species_equation_hydro}
\end{equation}

\begin{equation}
	\diff{}{t} \int_{V_\mathrm{f}}^{} \rho_\mathrm{f}\textbf{v}  \dint V_\mathrm{f} = -\int_{S_1}^{}p_\mathrm{f}\hat{\textbf{n}} \dint S_1 \,,
	\label{momentum_equation}
\end{equation}

\begin{equation}
	\diff{}{t} \int_{V_\mathrm{f}}^{} u_\mathrm{f}  \dint V_\mathrm{f} = -\int_{V_\mathrm{f}}^{}p_\mathrm{f}\left(\nabla\cdot\textbf{v}\right)\mathrm{d}V_\mathrm{f} \,.
	\label{energy_equation}
\end{equation}

\noindent Introducing the volume-averaged quantities for the mass density $\rho$, species number densities $n_j$, momentum density $\rho \textbf{v}$ and internal energy density $u$, and assuming a uniform pressure distribution\footnote{We are omitting the macron for the average to avoid overcomplicating the mathematical notation.},

\begin{equation}
	\diff{\left(\rho_\mathrm{f} V_\mathrm{f}\right)}{t} = 0 \,,
	\label{continuity_equation_uniform}
\end{equation}

\begin{equation}
	\diff{\left(n_j V_\mathrm{f}\right)}{t} = 0 \quad \mathrm{for}~j=0,\dots, N_\mathrm{S}\,,
	\label{species_equation_uniform}
\end{equation}

\begin{equation}
	\diff{\left(\rho_\mathrm{f} \textbf{v} V_\mathrm{f}\right)}{t} = -p_\mathrm{f}\int_{S_1}^{}\hat{\textbf{n}}\dint S_1=\textbf{0} \,,
	\label{momentum_equation_uniform}
\end{equation}

\begin{equation}
	\diff{\left(u_\mathrm{f} V_\mathrm{f}\right)}{t} =-p_\mathrm{f}\int_{S_1}^{}\textbf{v}\cdot\hat{\textbf{n}} ~ \mathrm{d}S_1= -p_\mathrm{f}v_1S_1 \,,
	\label{energy_equation_uniform}
\end{equation}

\noindent where the global momentum equation becomes trivial under spherical  symmetry. Noticing that, for a sphere, $\dint V/\dint t=vS$, one finally obtains the equations for the volume-averaged mass density, species number densities and internal energy density of the fuel:

\begin{equation}
	\diff{\rho_\mathrm{f}}{t} = -\rho_\mathrm{f} v_1 \frac{S_1}{V_\mathrm{f}}
	\label{continuity_equation_final} \,,
\end{equation}

\begin{equation}
	\diff{n_j}{t} = -n_j v_1 \frac{S_1}{V_\mathrm{f}}
	\label{species_equation_final} \quad \mathrm{for}~j=0,\dots, N_\mathrm{S}\,,
\end{equation}

\begin{equation}
	\diff{u_\mathrm{f}}{t} = -\left(p_\mathrm{f}+u_\mathrm{f}\right)v_1\frac{S_1}{V_\mathrm{f}}
	\label{energy_equation_final} \,,
\end{equation}

\noindent where the additional term $u_\mathrm{f}v_1 \left(S_1/V_\mathrm{f}\right)$ arises because we are solving for a quantity per unit volume \cite{atzeni19862}. The energy equation is separately solved for each fluid component (ion, electron, radiation)\footnote{The subscript ``f" for the fuel individual components is dropped to simplify the mathematical notation. Since the pusher has a single temperature $T_{\mathrm{p}}$, this is unambiguous.},

\begin{equation}
	\diff{u_{k}}{t} = -\left(p_{k}+u_{k}\right)v_1\frac{S_1}{V_\mathrm{f}} \quad \mathrm{for}~k \in \{\mathrm{i, e, r}\}
	\label{energy_equation_component_final} \,,
\end{equation}

\noindent whose sum provides Equation \ref{energy_equation_final}. The same exercise can be applied (without $n_j$) to the spherical shell volume $V_\mathrm{p}$ representing the pusher region (Figure \ref{pic:main_quantities_and_geometry}), leading to the following equations:

\begin{equation}
	\diff{\rho_\mathrm{p}}{t} = -\rho_\mathrm{p} \frac{v_2S_2 - v_1 S_1}{V_\mathrm{p}}
	\label{pusher_continuity_equation_final} \,,
\end{equation}

\begin{equation}
	\diff{u_\mathrm{p}}{t} = -\left(p_\mathrm{p}+u_\mathrm{p}\right)\frac{v_2S_2-v_1S_1}{V_\mathrm{p}}
	\label{pusher_energy_equation_final} \,.
\end{equation}

\noindent This set of equations must be augmented with kinematic equations for the radii $R_1, R_2$,

\begin{equation}
	\diff{R_k}{t}=v_k \quad \mathrm{for} ~ k \in \{1,2\} \,,
	\label{radii_equations}
\end{equation}

\noindent and two equations of state to link the mass and internal energy densities to the thermodynamic pressure for the two materials,

\begin{equation}
	\mathcal{F}_k\left(\rho_k, u_k, p_k\right)=0 \quad \mathrm{for} ~ k \in \{\mathrm{f, p}\} \,.
	\label{equation:eos}
\end{equation}

\noindent Our 2T EoS is based on the Frankfurt EoS (\texttt{FEOS}) package   \cite{Faik_ComputPhysCommun_2018}, which derives from the Quotidian EoS (\texttt{QEOS}) framework \cite{More_PhysFluids_1988} and is tabulated for temperatures up to $T_\mathrm{i}=100~\unit{keV}$.
Finally, one has to include equations for the surface velocities $v_1, v_2$ to close the system. Depending on how the equation for the implosion velocity is derived, three different hydrodynamic models can be defined, which determine the dynamics of $v_1, v_2$.
 
\subsubsection{Hydrodynamic model 0}
\noindent This is one of the most widely used hydrodynamics treatments for reduced ignition models \cite{Huang_PhysPlasmas_2017, Dodd_PhysPlasmas_2020}, in which an incompressible pusher with finite mass and an infinitesimally small thickness is assumed ($\Delta_\mathrm{p}\rightarrow0$). The fuel kinetic energy is considered to be zero. Therefore, the only force decelerating the pusher derives from the fuel pressure exerted on the surface $S_1$,

\begin{equation}
	m_\mathrm{p}\diff{v_1}{t} = p_\mathrm{f} S_1 \,.
	\label{velocity_equation_no_pusher}
\end{equation}

\noindent This model provides no information on the state of the pusher during the implosion phase, assuming its internal energy is constant. This poses serious problems in terms of energy distribution in the system, since geometry convergence is essentially neglected. 

We report this model for the sake of completeness, since its  inadequacy has been considered in previous works \cite{betti2001hot, betti2002deceleration} in the context of hot-spot ignition, and it will not be investigated any further.

\subsubsection{Hydrodynamic model I}
\noindent In this model we remove the assumption of an incompressible pusher with zero thickness, introducing a compressible pusher with a finite volume and a spatially uniform velocity profile, which thus has a \textit{constant thickness}. The initial work of Kirkpatrick and Wheeler adopted this approach \cite{ Kirkpatrick_NuclFusion_1975, Kirkpatrick_NuclFusion_1979}, limiting however the analysis to thin shells. In this case $v_1=v_2$, and an equation for the interface velocity $v_1$ can be obtained imposing the conservation of the total energy $E_{\mathrm{tot}}$ (assuming a closed thermodynamic system),

\begin{equation}
	\diff{E_{\mathrm{tot}}}{t}=\diff{\left(K_\mathrm{p}+U_\mathrm{p}+K_\mathrm{f}+U_\mathrm{f}\right)}{t}=0 \,,
	\label{global_energy_conservation}
\end{equation}

\noindent where the total kinetic and internal energies are respectively defined as $K_i=k_i V_i$ and $U_i=u_i V_i$ for $i \in \{\mathrm{f, p}\}$. The assumption of flat velocity profile provides $\dint K_\mathrm{p}/\dint t = m_\mathrm{p}v_1\left(\dint v_1/\dint t\right)$, whilst from Equations \ref{energy_equation_final}, \ref{pusher_energy_equation_final} we can write $\dint U_\mathrm{p}/\dint t=-p_\mathrm{p}v_1\left(S_2-S_1\right)$ and  $\dint U_\mathrm{f}/\dint t=-p_\mathrm{f}v_1S_1$. Following Kirkpatrick's approach of assuming a linear velocity profile in the fuel region\footnote{This is also substantiated analysing 1D hydrodynamic simulations carried out with our in-house hydrodynamic  code \texttt{B2} (Section \ref{section:comparison_b2}). Specifically, assuming a profile $v\left(r\right)=v_1\frac{r}{R_1}$, the fuel kinetic energy can be written as $K_\mathrm{f}=\frac{1}{2}\rho_\mathrm{f}\int_{0}^{R_1}4\pi r^2v^2\left(r\right)\dint r$.}, we can write $\dint K_\mathrm{f}/\dint t=\left(3/5\right)m_\mathrm{f}v_1\left(\dint v_1/\dint t\right)$. Therefore, we obtain 

\begin{equation}
	\left(m_\mathrm{p}+\frac{3}{5}m_\mathrm{f}\right)\diff{v_1}{t}=p_\mathrm{f}S_1 + p_\mathrm{p}\left(S_2-S_1\right) \,,
	\label{velocity_equation_const_thickness_pusher}
\end{equation}

\noindent which reverts back to Equation \ref{velocity_equation_no_pusher} if $S_2=S_1$ (and neglecting $K_\mathrm{f}$), i.e., when the pusher has a vanishing volume.

\subsubsection{Hydrodynamic model II}
\label{subsubsection:lagrangian_hydro}
\noindent Moving away from the constant-thickness pusher assumption can be challenging, since the only physical arguments  available to link the two velocities $v_1, v_2$ is momentum conservation, which we initially discarded. The strategy followed in work is to derive a non-uniform velocity profile in the pusher is to discretise the whole target in $N_\mathrm{c}=N_{\mathrm{c, f}} + N_{\mathrm{c, p}}$ (fuel + pusher)  concentric fluid elements (cells) whose masses are constant (in time) and comoving with the fluid. This is a Lagrangian representation of the hydrodynamics, which allows one to efficiently model situations in which the material strongly compresses or expands.  This representation is particularly advantageous in 1D and, mathematically, allows one to ignore the advection term in the transport equations. For each fluid element $j$:

\begin{figure}
	\centering
	\subfloat[]
	{\includegraphics[width=.47\textwidth,height=0.24\textheight]{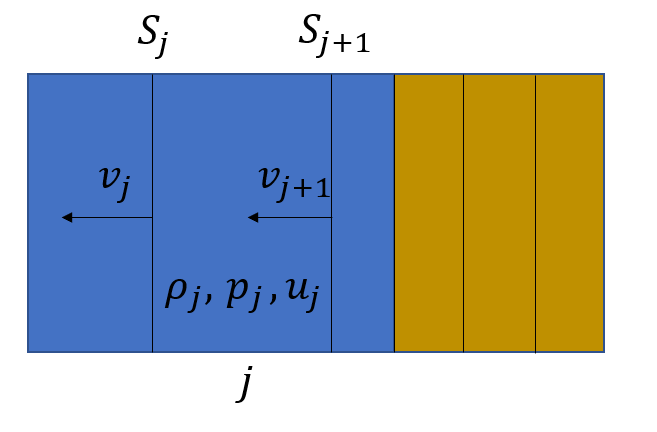}} \quad
	\subfloat[]
	{\includegraphics[width=.47\textwidth,height=0.24\textheight]{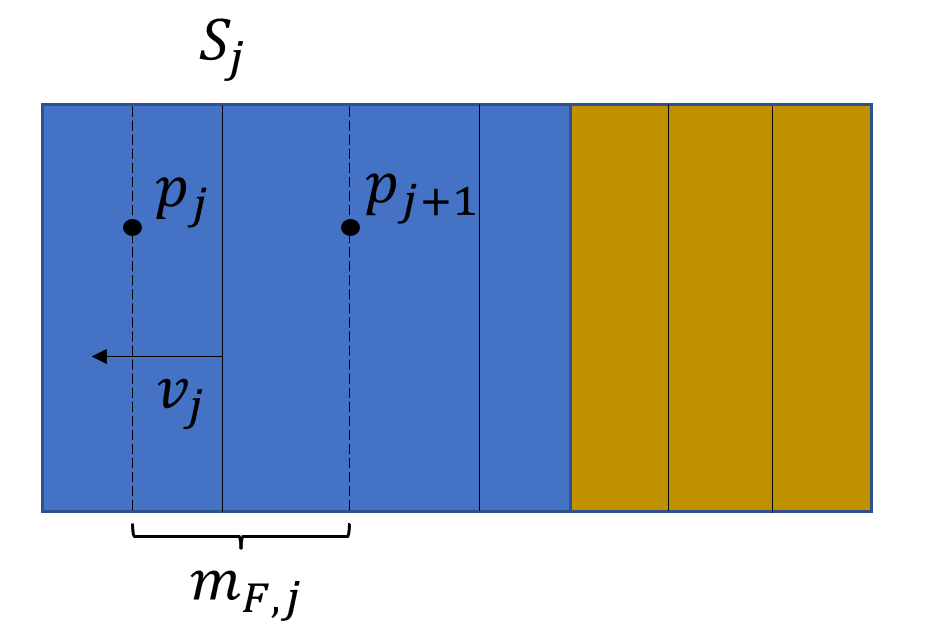}}
	\caption{Discretisation of fuel and pusher region using a staggered grid (a). Physical quantities necessary to define the momentum equation (Equation \ref{momentum_equation_lagrangian}) (b).}
	\label{pic:lagrangian_grids}
\end{figure}

\begin{equation}
	\diff{\rho_j}{t}=-\frac{\rho_j}{V_j}\left(v_{j+1}S_{j+1}-v_jS_j\right) \,,
	\label{density_equation_lagrangian}
\end{equation}

\begin{equation}
	\diff{u_{k, j}}{t} = -\left(p_{k, j}+u_{k, j}\right)\frac{v_{j+1}S_{j+1}-v_{j}S_{j}}{V_j} \quad \mathrm{for}~k \in \{\mathrm{i, e, r}\} \,.
	\label{energy_equation_lagrangian}
\end{equation}

\noindent We adopt a staggered grid approach (Figure \ref{pic:lagrangian_grids}), in which all quantities are stored at the cell centres, except for velocities, which are stored at face centres. The grid motion is provided by Equation \ref{radii_equations} (for each face, i.e., $k=j$), whilst the fluid elements interact via momentum transfer,

\begin{equation}
	\diff{v_j}{t}=\frac{p_j-p_{j+1}}{m_{F, j}}S_j \,,
	\label{momentum_equation_lagrangian}
\end{equation}

\noindent in which a projected face-centred mass $m_{F, j}$ represents the inertia of face $j$ subjected to the local pressure difference $p_j-p_{j+1}$ (Figure \ref{pic:lagrangian_grids}b). The boundary conditions are simply $v_0=0$ (fuel centre) and $p_{\mathrm{N_c+1}}=0$ (enforced by a ghost cell $j=N_{\mathrm{c}}+1$ at the outer boundary). Some form of artificial viscosity is necessary to capture shocks and avoid unphysical oscillations in the vicinity of  hydrodynamic discontinuities. The basic form from Von Neumann et al. \cite{vonneumann1950method} is adopted, in which a pseudo-pressure $p^*$, defined as  

\begin{equation}
	p^*=\rho c_q \Delta x^2 \left(\diff{v}{x}\right)^2 \quad \mathrm{if} ~ \diff{v}{x} < 0 \\,
	\label{artificial_viscosity}
\end{equation}

\begin{table}
	\centering
	\begin{tabular}{lll}
		\toprule
		Name & Description & Equations \\
		\midrule
		Hydro 0 & \makecell[l]{Volume averaged 0D model.\\ Pusher thickness is zero. \\ $v_\mathrm{p}\left(t, r\right)=0$.} &
		\makecell[l]{\ref{continuity_equation_final}, \ref{species_equation_final}, \ref{energy_equation_component_final}, \\
			\ref{radii_equations}, \ref{velocity_equation_no_pusher}}  \\
		\midrule
		Hydro I & \makecell[l]{Volume averaged 0D model.\\ Pusher thickness is constant. \\ $v_\mathrm{p}\left(t, r\right)=v_1(t)=v_2(t)$. \\
			$v_\mathrm{p}(t, r)$ derived from global energy conservation.} &
		\makecell[l]{\ref{continuity_equation_final},
			\ref{species_equation_final}, \ref{energy_equation_component_final},
			\ref{pusher_continuity_equation_final}, \\ \ref{pusher_energy_equation_final},
			\ref{radii_equations}, \ref{velocity_equation_const_thickness_pusher}}  \\
		\midrule
		Hydro II & \makecell[l]{1D Lagrangian model with artificial viscosity. \\ Pusher thickness is variable. \\ $v_\mathrm{p}(t, r)=\mathcal{F}(t, r)$. \\ $v_\mathrm{p}(t, r)$ derived from momentum conservation.} & \makecell[l]{
			\ref{species_equation_final},
			\ref{density_equation_lagrangian}, \ref{energy_equation_lagrangian}, \\ \ref{momentum_equation_lagrangian}, \ref{artificial_viscosity}} \\
		\bottomrule
	\end{tabular}
	\caption{Summary of the hydrodynamic models investigated in this work and corresponding equations.}
	\label{table:hydro_models}
\end{table}

\noindent is added to the thermodynamic pressure $p$ (in Equations \ref{energy_equation_lagrangian}, \ref{momentum_equation_lagrangian}) if the flow is compressive. Artificial viscosity is responsible for the entropy production, and its dissipative nature is a direct result of the requirement for the shock to have a finite width in terms of grid spacing $\Delta x$ (the constant is chosen to be $c_q=2$, corresponding to $\sim 3-4$ cell thick shocks \cite{mattsson2015artificial}).

Since the remaining operators in \texttt{FLAIM} evolve integrated conserved quantities, it is necessary to apply the following average after the Lagrangian hydrodynamic step:

\begin{equation}
	\bar{y}_k =\frac{\sum_{j}^{N_{\mathrm{c}, k}}y_{k, j}V_j}{\sum_{j}^{N_{\mathrm{c}, k}}V_j} \quad \mathrm{for}~y \in \{\rho, u\} ~ \mathrm{and}~k \in \{\mathrm{f, p}\} \,.
\end{equation}

\noindent The fuel and pusher  kinetic energies are computed as:

\begin{equation}
	K_k =\frac{1}{2}\sum_{j}^{N_{\mathrm{c}, k}+1}m_{F, j}v_j^2 \quad \mathrm{for}~k \in \{\mathrm{f, p}\} \,,
\end{equation}

\noindent with no need to hypothesise a velocity profile for the fuel as done for the Hydro I model. Finally, the surface velocities are simply $v_1=v_{N_{\mathrm{c, f}}+1}$ and $v_2=v_{N_{\mathrm{c}}+1}$.

\subsection{Thermal equilibration}
The differential heating resulting from the operators described above leads to temperature separation between the electrons and ions (a single temperature is assumed for all ion species) in the fuel, which will be equilibrated over time through energy exchange due to inter-particle collisions and interactions with the mean field. For the single-fluid description adopted in \texttt{FLAIM} (as well as in our hydrodynamics codes) the rates of change of the internal energy densities are given by

\begin{equation}
	\diff{u_{\mathrm{i}}}{t}=\dot{Q}_{\mathrm{ie}} \,,
	\label{thermal_eq_i}
\end{equation}

\begin{equation}
	\diff{u_{\mathrm{e}}}{t}=\dot{Q}_{\mathrm{ei}} \,.
	\label{thermal_eq_el}
\end{equation}

\noindent The volumetric energy exchange rates, $\dot{Q}_{\mathrm{ie}} = -\dot{Q}_{\mathrm{ei}}$, can be evaluated in-line or by interpolation on pre-computed tables based on a variety of models. The standard model used in \texttt{FLAIM} is the reduced Fermi golden rule (RFGR) approach \cite{Hazak_PhysRevE_2001, Gericke_JPhysConfSeries_2005}, which compares well to Molecular Dynamics (MD) simulations for simple materials such as DT under the conditions of interest \cite{Murillo_PhysRevLett_2008, Vorberger_HEDP_2014}. Non-ideal effects such as large-angle binary collisions \cite{Gericke_PhysRevE_2002}, local field corrections \cite{Daligault_PhysRevE_2009, Daligault_PhysRevE_2007} and the coupled mode effect \cite{Chapman_PhysRevE_2013, Dharma-wardana_PhysRevE_1998, Vorberger_PhysPlasmas_2009, Vorberger_PhysRevE_2010} are not important for the conditions achieved in burning fusion plasmas.

\subsection{Wall}
\noindent Heat losses from the fuel to the pusher are mediated by the presence of a \textit{wall} (i.e., the fuel-pusher interface). The temperature difference between the fuel and the wall governs the heat losses, namely thermal conduction and radiation loss. The temperature of the wall is determined by a steady-state balance of the heat fluxes $F\left(t\right)$ from the fuel (radiation, conduction and $\alpha$ particles) and the flux diffused into the pusher,

\begin{equation}
	F\left(t\right)=-\frac{4}{3}\frac{1}{\chi}\frac{\partial \sigma T_\mathrm{p}^4}{\partial x}\bigg|_{x=0} \,,
	\label{balance_of_fluxes}
\end{equation}

\noindent where $\sigma$ is the Stefan-Boltzmann constant and  $\chi$ is the material opacity. The flux on the right-hand side is transported into the pusher by diffusion assuming local supersonic conditions (i.e., no convection). Therefore, the temperature gradient in Equation \ref{balance_of_fluxes} is provided by the 1D planar diffusion equation

\begin{equation}
	\frac{\partial u_\mathrm{p}}{\partial t}=\frac{4}{3}\frac{\partial}{\partial x}\left(\frac{1}{\chi}\frac{\partial \sigma T_\mathrm{p}^4}{\partial x}\right) \,.
	\label{radiation_diffusion}
\end{equation}

\noindent The solution of Equation \ref{radiation_diffusion} (using Equation \ref{balance_of_fluxes} as time-varying boundary condition and an appropriate EoS for the pusher, $\mathcal{F}_\mathrm{p}\left(\rho_\mathrm{p}, u_\mathrm{p}, p_\mathrm{p}\right)=0$) provides the wall temperature:

\begin{equation}
	T_\mathrm{w}\left(t\right)=T_\mathrm{p}\left(x, t\right)\vert_{x=0} \,.
	\label{equation_wall_temperature}
\end{equation}
 
\noindent Since \texttt{FLAIM} is a reduced model, we seek a simplified approach to the numerical solution of Equation \ref{radiation_diffusion}, that possibly removes any information of the spatial coordinates.  The works of Hammer et al. \cite{Hammer_PhysPlasmas_2003} and Dodd et al. \cite{Dodd_PhysPlasmas_2020}, present\deleted[id=P10]{s} an approximate approach for  deriving the wall temperature $T_\mathrm{w}$ and the position of the heat front $x_\mathrm{F}$ (with respect to the wall) in terms of simple time-dependent ODEs, assuming a infinitely steep front for the advancing heat wave, and a power-law form for the pusher EoS. \added[id=P10]{In our work, the coefficients for the power-law EoS are calculated by a least-squares fitting over a prescribed region of the tabulated EoS for the pusher material}. The pusher opacity is calculated using the the \texttt{SpK} code \cite{fraser2024}. We \replaced[id=P10]{direct readers}{refer} to \replaced[id=P10]{the reference}{those} works for more details about the derivation,  implementation and verification of the model.

\subsection{Thermal conduction}
\noindent Thermal conduction is accounted for following

\begin{equation}
	\diff{u_{k}}{t}=-\dot{Q}_{k, \mathrm{cond}} \,,
\end{equation}

\noindent where the conductive power density term is approximated as

\begin{equation}
	\begin{split}
		\dot{Q}_{k, \mathrm{cond}} &= \\
		&= \frac{1}{V_\mathrm{f}}\int_{V_\mathrm{f}}^{}\nabla\cdot\left(\kappa_k\nabla T_k\right)\dint V_\mathrm{f} \\
		&=  \frac{1}{V_\mathrm{f}}\int_{S_1}\kappa_k\nabla T_k \cdot \hat{\textbf{n}}\dint S_1 \approxeq  \kappa_k\frac{T_{k}-T_\mathrm{w}}{R_1} \frac{S_1}{V_\mathrm{f}}
	\end{split}
	\label{thermal_conduction_form}
\end{equation}

\noindent for $k \in \{\mathrm{i, e}\}$, accounting for ion and electron conduction, and where a linear temperature profile is assumed in the fuel. A flux-limiter based on the particle thermal speed $v_{\mathrm{th}}$ is applied to avoid the inadequacy of the classic Fourier form of the thermal flux in the vicinity of steep spatial gradients \cite{malone1975indications},

\begin{equation}
	\dot{Q}_{k, \mathrm{cond}} = \min\left(\dot{Q}_{k, \mathrm{cond}}, \beta_k v_{\mathrm{th}, k}u_k \frac{S_1}{V_\mathrm{f}}\right) \,,
	\label{flux_limited_conduction}
\end{equation}

\noindent where the coefficient $\beta_k=0.5, 0.05$ for ions and electrons respectively.  Work is presently underway to develop a reduced kinetic description of the heat flux due to ions \cite{Mitchell_PlasmaPhysControlFusion_2024}, the effects of which we will examine in a dedicated forthcoming publication. The total conductive power density is finally applied to the pusher to enforce energy conservation, 

\begin{equation}
	\diff{u_{\mathrm{p}}}{t}=\left(\dot{Q}_{\mathrm{i, cond}}+\dot{Q}_{\mathrm{e, cond}}\right) \frac{V_\mathrm{f}}{V_\mathrm{p}} \,.
\end{equation}

\noindent For the thermal conductivity of the ions, we use Stanton and Murillo's result derived from the perturbative solution to the Boltzmann transport equation \cite{Stanton_PhysRevE_2016}. For the electrons, we use a modified version of Lee and More's simple relaxation time framework \cite{Lee_PhysFluids_1984} corrected to account for electron-neutral and electron-electron collisions and suitable restrictions to the mean free path in the warm dense matter and solid phases. In both of the conductivities, the appropriate effective ion charges are taken from the work of Hoffman et al. \cite{Hoffman_PhysPlasmas_2015} for generalising the single-fluid framework to multi-component mixtures.

\subsection{Radiation loss}
\noindent The main mechanisms of electron-radiation interaction in the fuel are bremsstrahlung, Compton scattering and their inverse processes \cite{Huang_PhysPlasmas_2017, fraley1974thermonuclear, woodward1970compton}:

\begin{equation}
	\dot{Q}_{\mathrm{brem}} = \nu_c n_\mathrm{e} \frac{4}{\pi^{3/2}}Z_{\mathrm{eff}}\frac{k_Ce^2}{\hbar c}\sqrt{\frac{k_BT_{\mathrm{e}}m_\mathrm{e}c^2}{2}}I_B\left(\frac{T_{\mathrm{r}}}{T_{\mathrm{e}}}\right) \,,
	\label{bremsstrahlung}
\end{equation}

\begin{equation}
	\dot{Q}_{\mathrm{scatt}}=4\nu_cu_\mathrm{r}k_B\frac{T_\mathrm{e}-T_\mathrm{r}}{m_\mathrm{e} c^2} \,,
\end{equation}

\noindent where $k_B$ and $k_C$ define the Boltzmann and Coulomb constants respectively, $\nu_c$ is the basic frequency for Compton scattering and $I_B\left(x\right)=2R\left(x\right)$, with $R\left(x\right)$ defining the regular Hurwitz integral \cite{Molvig_PhysPlasmas_2009}. These sources set the radiation energy density $u_\mathrm{r}$ in the fuel cavity. The expression for $\dot{Q}_{\mathrm{brem}}$ is augmented accounting for the fuel optical-depth following the work from Dodd et al. \cite{Dodd_PhysPlasmas_2020}. The radiation flux leaving a spherical cavity  containing a radiative energy density $u_\mathrm{r}$ is $u_\mathrm{r}c/4$, whilst the radiation flux emitted from the wall to the fuel (assuming a black-body) is $\sigma T_\mathrm{w}^4$. The net radiative flux leaving the fuel volume is therefore:

\begin{equation}
	\dot{Q}_{\mathrm{rad}}=\left(u_\mathrm{r}\frac{c}{4}-\sigma T_\mathrm{w}^4\right)\frac{S_1}{V_\mathrm{f}} \,.
	\label{radiation_loss}
\end{equation}

\noindent These source terms are applied to the relevant energy densities as

\begin{equation}
	\diff{u_{\mathrm{e}}}{t}= - \dot{Q}_{\mathrm{brem}} - \dot{Q}_{\mathrm{scatt}} \,,
\end{equation}

\begin{equation}
	\diff{u_\mathrm{r}}{t}= \dot{Q}_{\mathrm{brem}} + \dot{Q}_{\mathrm{scatt}} -\dot{Q}_{\mathrm{rad}} \,.
\end{equation}

\noindent Finally, enforcing energy conservation for the whole system,
\begin{equation}
	\diff{u_\mathrm{p}}{t}=\dot{Q}_{\mathrm{rad}}\frac{V_\mathrm{f}}{V_\mathrm{p}} \,.
\end{equation}

\subsection{Thermonuclear burn}
\noindent The thermonuclear burn includes the main branches of DT/DD reactions, \deleted[id=P11]{alpha particle escape and fuel depletion.}

\begin{equation}
	\mathrm{D + T \rightarrow \underset{3.5~MeV}{\alpha} + \underset{14.1~MeV}n} \,,
\end{equation}	

\begin{equation}
	\mathrm{D + D \rightarrow \underset{1.01~MeV}{T} + \underset{3.03~ MeV}{p}} \,,
\end{equation}

\begin{equation}
	\mathrm{D + D \rightarrow \underset{0.82~MeV}{\prescript{3}{}{He}} + \underset{2.45~MeV}{n}} \,.
\end{equation}

\noindent In a general reactive system including $N_\mathrm{S}$ species \added[id=P11]{(6 in this case)} and $N_\mathrm{R}$ reactions \added[id=P11]{(3 in this case)}, one has to solve for the number density of each species $j$,

\begin{equation}
	\diff{n_j}{t} = s_j \sum_{k}^{N_\mathrm{R}}\mathcal{R}_k\nu_{j, k} \quad \mathrm{for}~j=0,\dots, N_\mathrm{S}\,,
	\label{species_equation}
\end{equation}

\noindent where $s_j$ is the fraction of species $j$ stopped in the fuel (if $j$ is a  product, otherwise $s_j=1$). For the $\alpha$ particles, $s_{\alpha}$ is provided by an alpha escape model such as Krokhin and Rozanov \cite{krokhin1973escape}. For neutrons, we assume complete escape $s_\mathrm{n}=0$.  $\nu_{j, k}$ is the stoichiometric coefficient of species $j$ in reaction $k$. The reaction rates $\mathcal{R}_k$ are based on the Bosch-Hale fits for the reactivities \cite{Bosch_NuclFusion_1992}, with the option to use  tabulated ones to account for the reactivity reduction due to Knudsen loss \cite{Molvig_PhysRevLett_2012, Albright_PhysPlasmas_2013}. The fuel internal energy density obeys

\begin{equation}
	\diff{u_\mathrm{f}}{t} = \sum_{i}^{N_\mathrm{R}}\mathcal{R}_i\sum_{j}^{N_{\mathrm{P}, i}} K_{j, i}s_j \,,
	\label{fuel_internal_energy_burn_equation}
\end{equation}

\noindent in which $K_{j, i}$ represents the kinetic energy of the product $j$ in reaction $i$, scaled by the relative stopping fraction $s_j$, which is summed over the $N_{\mathrm{P}, i}$ products of reaction $i$. This rate is subsequently split into ion and electron contributions based  on established models, such as those of Fraley et al. \cite{Fraley_PhysFluids_1974} or Atzeni and Caruso \cite{Atzeni_book}. The fraction of each species $j$ escaping the fuel is assumed to be entirely stopped by the pusher,

\begin{equation}
	\diff{u_\mathrm{p}}{t} = \sum_{i}^{N_\mathrm{R}}\mathcal{R}_i\sum_{j}^{N_{\mathrm{P}, i}} K_{j, i}\left(1-s_j\right) \,.
	\label{pusher_internal_energy_burn_equation}
\end{equation}

\noindent Finally, the fuel mass density is re-evaluated based on the updated number densities $n_j$ and the masses of the ions $m_j$, 

\begin{equation}
	\rho_\mathrm{f}=\sum_{j}^{N_\mathrm{S}}n_jm_j \,,
\end{equation}

\noindent from which fuel depletion naturally follows from the fact that $n_\mathrm{n}$ (neutron number density) is always null (since $s_\mathrm{n}=0$).

\section{Implementation}
\label{section:numerical_setup}
\noindent In this section we provide a brief description of how the model is numerically solved and an overview of the code structure and implementation. 

\subsection{Numerical integration}
\noindent The set of ODEs defining the mathematical model are solved with an operator-splitting approach, in order to efficiently handle the wide range of characteristic time-scales the different physical phenomena evolve on (e.g., very fast burn and relatively slow hydrodynamics). Given $N_\mathrm{o}$ physical operators (i.e., hydro, thermal conduction, radiation loss etc.), we can write

\begin{equation}
	\textbf{y}' = \textbf{f}\left(\textbf{y}', \textbf{y}\right) = \sum_{k}^{N_\mathrm{o}}\textbf{f}_k\left(\textbf{y}', \textbf{y} \right) \,.
\end{equation}

\noindent The numerical integration for each operator is based on a first-order explicit forward Euler, obtaining the following sequence for each time step $\Delta t_0$:

\begin{equation}
	\begin{cases}
		\textbf{y}_{1}^{n+1} = \textbf{y}_{0}^n + \textbf{f}_0\left(\textbf{y}'^n_0, \textbf{y}^n_0\right)\Delta t_0 \,, \\
		\textbf{y}_{2}^{n+1} = \textbf{y}_{1}^{n+1} + \textbf{f}_1\left(\textbf{y}'^{n+1}_1, \textbf{y}^{n+1}_1\right)\Delta t_0 \,, \\
		\vdots \\
		\textbf{y}_{N_\mathrm{o}}^{n+1} = \textbf{y}_{N_\mathrm{o}-1}^{n+1} + \textbf{f}_{N_\mathrm{o}-1}\left(\textbf{y}'^{n+1}_0, \textbf{y}^{n+1}_0\right)\Delta t_0 \,,
	\end{cases}
\label{operator_splitting}
\end{equation}

\noindent where the subscripts $i=0,\dots, N_\mathrm{o}$ identify the partial evolution of $\textbf{y}$ due to operator $i$. Equation \ref{operator_splitting} defines a Lie-Trotter splitting \cite{trotter1959product}, since it involves a single forward Euler step using $\Delta t_0$   for each operator. In \texttt{FLAIM} we also have the option of using a Strang splitting \cite{strang1968construction}, in which the non-hydrodynamic physical operators are advanced for $\Delta t_0/2$, before and after the hydrodynamic one. We use Lie-Trotter splitting for most \texttt{FLAIM} simulation due to the lower computational cost, with Strang splitting only used in cases where extra time accuracy is required (\ref{app:operator_splitting}).

The inherent conditional stability of explicit schemes is satisfied by a subcycling methodology. The first operator $\textbf{f}_0$ is the hydrodynamics, employing a time step $\Delta t_0$, and different time steps $\Delta t_k$ are used for the remaining $\textbf{f}_k$ operators, either based on their specific stability conditions or determined by imposing a specific requirement for the solution at the next step (e.g., setting a maximum fractional change of one or more variables). If $\Delta t_k < \Delta t_0$, the operator $\textbf{f}_k$ is internally evolved multiple times until $\sum_{j}^{}\Delta t_{j, k}=\Delta t_0$, otherwise  $\Delta t_0$ is used\footnote{The subscript $j$ for the internal time steps $\Delta t_{j, k}$ is justified by the fact that they can potentially change at every sub-step.}. The hydro time step $\Delta t_0$ is computed for the model Hydro I (Table \ref{table:hydro_models}) as 

\begin{equation}
	\Delta t_{0} = \eta \frac{R_1}{v_1} \,,
\end{equation}

\noindent with $\eta$ is a user-specified safety factor (usually ~$0.01\leq \eta \leq 0.001$). For the Hydro II model, $\Delta t_{0}$ is based on the classic Courant–Friedrichs–Lewy (CFL) condition

\begin{equation}
	\Delta t_{0} = a\min\left(\frac{\Delta x_j}{\vert v_j + c_{s, j}\vert}\right)  \quad  \mathrm{for} ~ j=0,\ldots, N_\mathrm{c} \,,
	\label{cfl_condition}
\end{equation}

\noindent where $v_i$ is the cell-centred (linearly interpolated) grid velocity and $c_{s, i}$ is the local sound speed, extracted from the EoS table. The Courant number $a$ is typically chosen to be  $0.1 \leq a \leq 0.5$.

\subsection{Code structure}
\noindent \texttt{FLAIM} is developed as a stand-alone library written entirely in \texttt{C++}.  The code core is characterised by a modular structure, in order to provide a versatile tool that allows developers to easily implement and switch between models with minimal effort. To this purpose, \texttt{FLAIM} heavily relies on the object-oriented features of the \texttt{C++} language for the implementation of the physical operators. This allows the users to easily implement various time integration methods or different versions of the same operator, and to test them rapidly without changing any aspect about the general code structure.

The \texttt{FLAIM} external API is written in Python, using the \texttt{Boost.Python} library to enable the interoperability between the two programming languages. This has multiple advantages, including a facilitated interaction with a user-friendly language, immediate post-processing capabilities, an extra layer of control on the functionality exposed to the API (in addition to the   encapsulation provided by \texttt{C++}), as well as the complete separation between interface and implementation.
\section{Comparison with 1D hydrodynamic simulations}
\label{section:comparison_b2}
\noindent To use \texttt{FLAIM} as  an efficient and  reliable surrogate for exploring   and for  optimising target designs, one must assess its accuracy in a wide region of the design space with respect to well-established high fidelity models. In this section we focus on the analysis of the hydrodynamic operator\footnote{The simulations showed in this Section were conducted with thermal equilibration for both codes (Equations \ref{thermal_eq_i}, \ref{thermal_eq_el}), to avoid the preferential heating of any component of the fluid.} (Table \ref{table:hydro_models}).

At FLF we have developed a suite of numerical capabilities to support our experimental activity and design work, including a Resistive-Magneto-Hydrodynamic (RMHD) code called \texttt{B2}. \replaced[id=P9]{We refer to \ref{app:b2_code} for a brief description of the \texttt{B2} code, as well as for a list of verification test cases for the principal operators/models relevant for this work.}{\texttt{B2} works on regular orthogonal 3D grids and encompasses (i) Eulerian  hydrodynamics with Lagrangian re-map and artificial viscosity; (ii) Volume Of Fluid (VOF) method for interface tracking \cite{hirt1981volume}, and a Simple Line Interface Calculation (SLIC) method for the local interface reconstruction; (iii)  multi-group Eulerian radiation transport ($\mathbb{P}_{1/3}$ closure); (iv) flux-limited ion and electron thermal conduction; (v) fusion burn, including fuel depletion and single-group Eulerian $\alpha$ particle transport for fusion self-heating \cite{atzeni1981diffusive}. 
\texttt{B2} is continuously and extensively verified against a suite of test-cases and validated with in-house experimental results. Cross-code comparison against equivalent hydro codes shows satisfactory agreement and it is currently our main production code for the design and optimisation of our advanced targets. In this work \texttt{B2} is used as a high-fidelity reference to evaluate the predictive capabilities of \texttt{FLAIM} and to assess its adequacy as a reduced model.} 

\subsection{Simulation scan}
\label{subsec:simulation_scan}
\noindent In order to quantify the physical accuracy of \texttt{FLAIM} against \texttt{B2}, a quantitative comparison over a large parameter space holds the utmost importance. We have run a scan of 125 1D spherical \texttt{B2} simulations that represent variations of the Revolver design point \cite{Molvig_PhysRevLett_2016}. The pusher material is gold ($\rho_\mathrm{p}=19.3~\unit{kg/m^3}$) and the fuel is an equimolar mixture of DT at $T_\mathrm{i}=T_{\mathrm{e}}=33~\unit{K}$, with an initial density of $\rho_\mathrm{f}=0.173~\unit{kg/m^3}$. The \texttt{B2} simulations are initialised with a flat velocity profile in the pusher and $p_{\mathrm{N_c + 1}}=0$ outside the capsule. Table \ref{table:b2_scan} summarises the parameters defining the scan.

\begin{table}
	\centering
	\begin{tabular}{llll}
		\toprule
		& \quad \quad Revolver \cite{Molvig_PhysRevLett_2016} & \quad \quad \texttt{B2} scan & n points \\
		\midrule
		$R_1$ $[\unit{\mu m}]$ & \quad \quad $326$ & \quad \quad $150-450$ & \quad \quad 5\\ 
		\midrule
		$\Delta_\mathrm{p}$ $[\unit{\mu m}]$ & \quad \quad $60$ & \quad \quad $30-100$ & \quad \quad 5\\ 
		\midrule
		$v$ $[\unit{km/s}]$ & \quad \quad $200$ & \quad \quad $100-300$ & \quad \quad 5\\ 
		\bottomrule
	\end{tabular}
	\caption{Parameter space explored by the \texttt{B2} simulation scan (125 simulations, for each coordinate the n points are equidistant) in terms of initial fuel radius $R_1$, initial pusher thickness $\Delta_\mathrm{p}$ and initial implosion velocity $v=v_1=v_2$. Revolver parameters are reported for reference. \texttt{B2} simulations are run with a resolution of $\sim0.6~\unit{\mu m/cell}$ based on a standard convergence analysis.}
	\label{table:b2_scan}
\end{table}

\texttt{FLAIM} and \texttt{B2} will be compared based on the profiles of volume-averaged conserved quantities. Whilst these are naturally captured in \texttt{FLAIM} (see Section \ref{section:mathematical_model}), they must be explicitly post-processed from 1D \texttt{B2} simulations. For a generic quantity $y_{\texttt{B2}}$, we have for the fuel

 \begin{equation}
 	\bar{y}_{\texttt{B2}, \mathrm{f}}\left(t\right)=\frac{3}{R_1^3}\int_{0}^{R_1} r^2 y_{\texttt{B2}}\left(t, r\right)\dint r \,,
 	\label{average_B2_profiles_fuel}
 \end{equation}
 
 \noindent whereas for the pusher we can write 
 
  \begin{equation}
 	\bar{y}_{\texttt{B2}, \mathrm{p}}\left(t\right)=\frac{3}{R_2^3-R_1^3}\int_{R_1}^{R_2} r^2 y_{\texttt{B2}}\left(t, r\right)\dint r \,,
 	\label{average_B2_profiles_pusher}
 \end{equation}

 \noindent where $y \in \{u, \rho\}$, i.e. the conserved quantities defining the system. The average profiles for temperature $\bar{T}_{\texttt{B2}, k}\left(t\right)$ and pressure $\bar{p}_{\texttt{B2}, k}\left(t\right)$ for $k\in \{\mathrm{f, p}\}$  are derived from the fuel and pusher equations of state, which are the same for the two codes. The comparison metric is an $L_2$ norm for the fuel ion temperature $T_\mathrm{i}\left(t\right)$ and mass density   $\rho_\mathrm{f}\left(t\right)$ temporal profiles, defined as 
 
 \begin{equation}
 	L_2\left(y\right) = \sqrt{\sum_{i}^{N_{\mathrm{steps}}}\left(\frac{y_{\texttt{FLAIM}}\left(t_i\right) - \bar{y}_{\texttt{B2}}\left(t_i\right)}{\bar{y}_{\texttt{B2}}\left(t_i\right)}\right)^2}  \quad\quad \mathrm{for} ~ y \in \{T_\mathrm{i}, \rho_\mathrm{f}\} \,,
 	\label{l2_norm}
 \end{equation}
 
 \noindent where $y$ is the temporal profile of the physical quantity of interest and $N_{\mathrm{steps}}$ is the number of time steps of the simulation. This norm measures the average distance between the profiles predicted by \texttt{FLAIM} and \texttt{B2} and provides a metric to quantify the agreement between the two codes. 

\subsection{Hydro I}
\label{subsec:uniform_pusher_velocity}
\noindent We report here a brief description of how the  scan defined in Table \ref{table:b2_scan} is run using the model Hydro I (Table \ref{table:hydro_models}), based on the assumption of a spatially uniform velocity profile in the pusher (Equation \ref{velocity_equation_const_thickness_pusher}). As reported in several works \cite{Molvig_PhysRevLett_2016, Keenan_PhysPlasmas_2020b}, the implosion dynamics can be split in two different phases: (i) a shock-heating phase, which lasts until the initial shock reaches the origin; (ii) a quasi-isentropic compression phase that brings the fuel state towards the ignition conditions. The fuel energy equation for Hydro I in \texttt{FLAIM} (Equation \ref{energy_equation_component_final}) describes an adiabatic compression of a generic volume $V_\mathrm{f}$. This means that with \texttt{FLAIM} we are only able to describe the \textit{second} phase of the implosion dynamics, due to the lack of shock description in volume-averaged hydrodynamics. Therefore, for each \texttt{B2} simulation:

\begin{enumerate}
	\item We identify the time $t_\mathrm{ws}$ (warm-start) at which the initial shock reaches the origin and reflects back intersecting the fuel-pusher interface\footnote{The actual definition of shock-heating phase given by Molvig \cite{Molvig_PhysRevLett_2016} limits the dynamics until the shock converges on the origin. We decided to extend it to allow the residual weak shocks to reverberate and adhere more stringently to the adiabatic compression requirement that \texttt{FLAIM} necessitates.}, and extract the fuel and pusher states (in terms of volume-averaged $\rho, u$ and the pusher kinetic energy $K_\mathrm{p}$) and geometry ($R_1, R_2, \Delta_\mathrm{p}$) at $t_\mathrm{ws}$;
	\item This intermediate configuration is used to initialise a \texttt{FLAIM} simulation and to run it until completion ($t=t_\mathrm{end}$);
	\item Using the metric defined in Equation \ref{l2_norm}, the \texttt{B2} and \texttt{FLAIM} outputs are compared  for $t_\mathrm{ws}\leq t \leq t_\mathrm{end}$.
\end{enumerate}

\subsection{Hydro II}
\noindent Using the model Hydro II, the condition of  quasi-isentropic transformation is not required: the fluid elements are still described by an quasi-isentropic transformation, but the presence of artificial viscosity allows us to track the shock propagation, as well as stabilising the numerical solution \cite{mattsson2015artificial}. This makes the model Hydro II suitable for the modelling of both phases of the implosion dynamics, allowing us, in principle, to compare the \texttt{FLAIM} and \texttt{B2} directly from cold conditions. However, to maintain consistency with the approach used for the model Hydro I and to provide a solid comparison based on \textit{equivalent} initial conditions, we performed the scan following the steps $1-3$ defined in Section \ref{subsec:uniform_pusher_velocity} (i.e., starting from $t_\mathrm{ws}$). The comparison is still based on the $L_2$ norm defined in Equation \ref{l2_norm}.

\subsection{Quantitative comparison: L2 norms}
\begin{figure}
	\centering
	{\includegraphics[width=1.\textwidth,height=0.43\textheight]{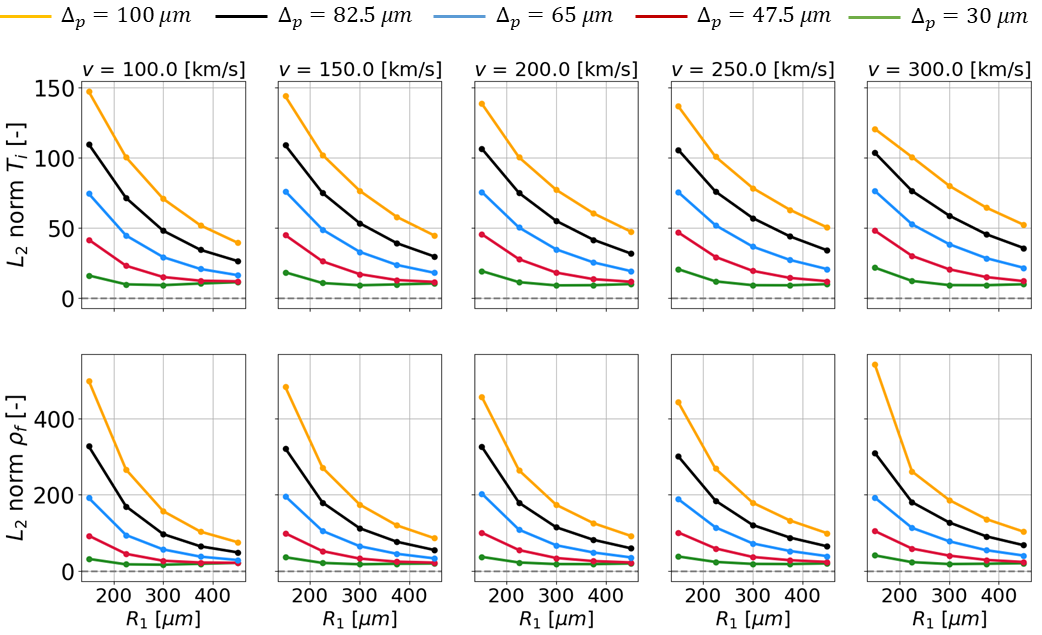}}
	\caption{$L_2$ norms (Equation \ref{l2_norm}) for the comparison \texttt{B2}-\texttt{FLAIM} for the fuel ion temperature $T_\mathrm{i}$ (top row) and mass density $\rho_\mathrm{f}$ (bottom row) profiles at different initial fuel radii $R_1$, pusher thicknesses $\Delta_\mathrm{p}$ and implosion velocities $v$. Hydro I model (Table \ref{table:hydro_models}).}
	\label{pic:l2_norms}
\end{figure}

\noindent Figure \ref{pic:l2_norms} reports the $L_2$ norms for ion temperature $T_\mathrm{i}$ and mass density $\rho_\mathrm{f}$ in the fuel for the scan using Hydro I, as a function of the initial fuel radius $R_1$ at different initial implosion velocities $v$ and pusher thicknesses $\Delta_\mathrm{p}$. One immediately notices a common behaviour of the $L_2$ norm in the parameter space: (i) it decreases with the initial fuel radius $R_1$, (ii) it increases with the initial pusher thickness $\Delta_\mathrm{p}$ and  (iii) it is weakly dependent on the initial implosion velocity $v$.

\begin{figure}
	\centering
	{\includegraphics[width=1.\textwidth,height=0.43\textheight]{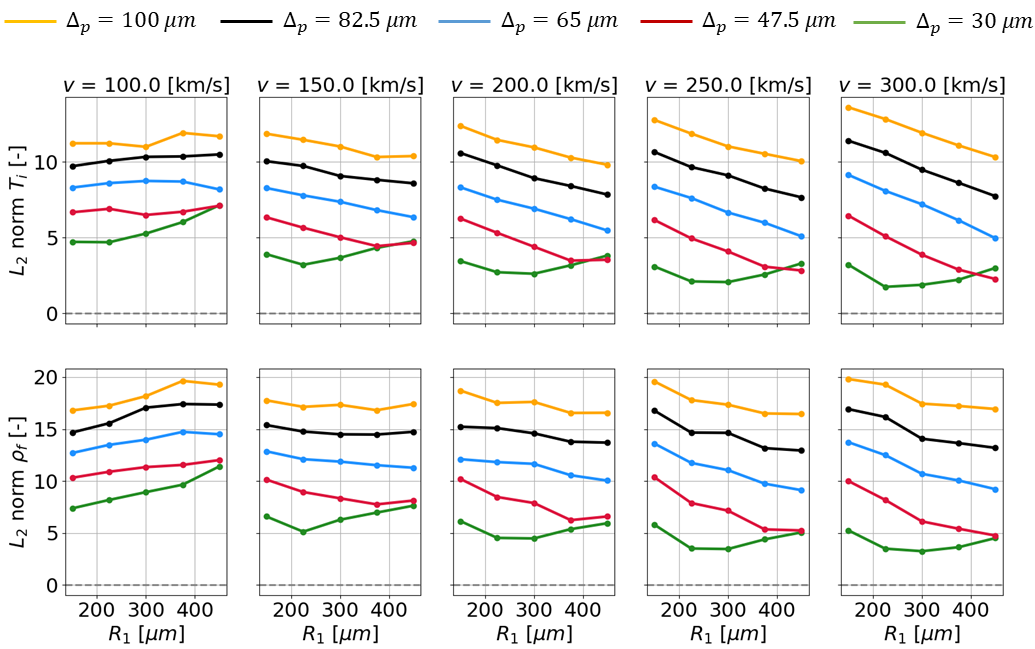}}
	\caption{$L_2$ norms (Equation \ref{l2_norm}) for the comparison \texttt{B2}-\texttt{FLAIM} for the fuel ion temperature $T_\mathrm{i}$ (top row) and mass density $\rho_\mathrm{f}$ (bottom row) profiles at different initial fuel radii $R_1$, pusher thicknesses $\Delta_\mathrm{p}$ and implosion velocities $v$. Hydro II model (Table \ref{table:hydro_models}).}
	\label{pic:l2_norms_lagrangian}
\end{figure}

\begin{figure}
	\centering
	{\includegraphics[width=0.8\textwidth,height=0.32\textheight]{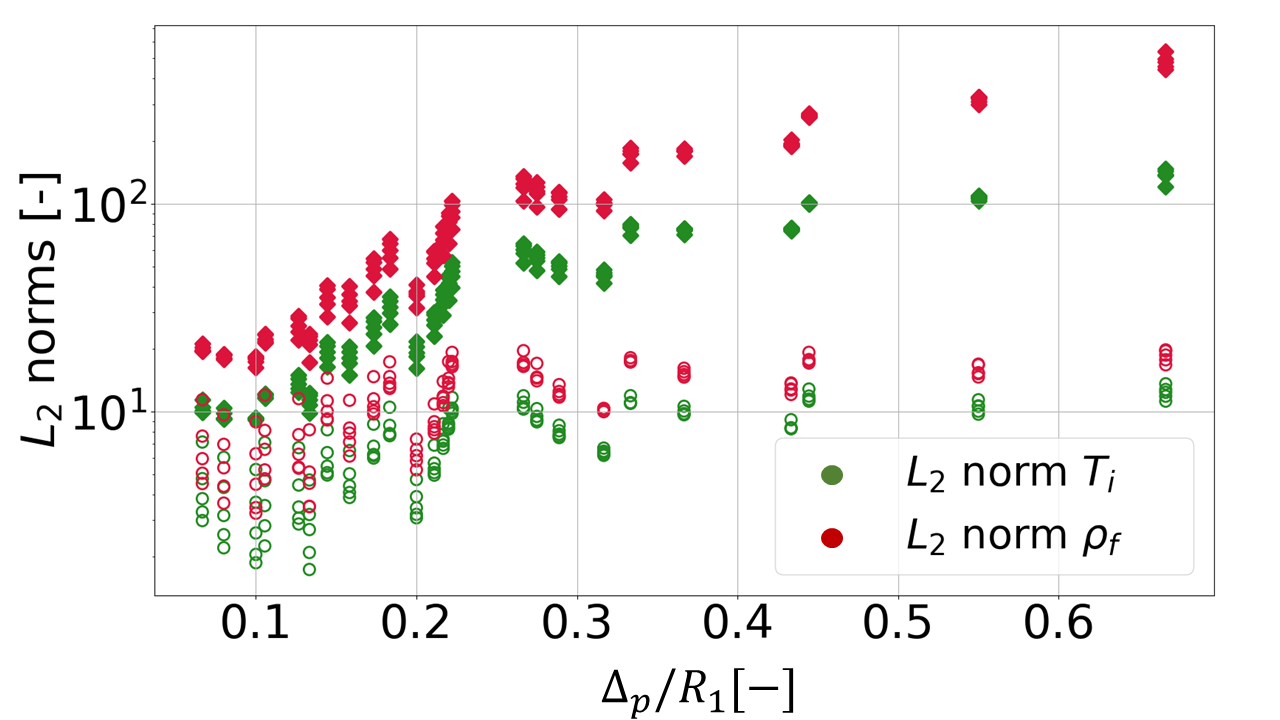}}
	\caption{Scatter plot of $L_2$ norms of the ion temperature $T_\mathrm{i}$ and fuel density $\rho_\mathrm{f}$ vs. $\Delta_\mathrm{p}/R_1$ for the scan defined in Table \ref{table:b2_scan}. Diamonds for the model Hydro I, empty circles for the model Hydro II.}
	\label{pic:l2_norm_scatterplot}
\end{figure}

In contrast, the $L_2$ norm profiles provided by Hydro II (Figure \ref{pic:l2_norms_lagrangian}), show significant differences. First, the average magnitude of the $L_2$ norm has considerably decreased (by a factor of $\sim10-40$), especially for the density profile, denoting a much better agreement between the profiles. Secondly, the strong variation with respect to $R_1$ and $\Delta_\mathrm{p}$ is no longer visible, and a very weak dependence is observed. This aspect is further exemplified by the scatter plot of the values of $L_2$ norms in function of the $\Delta_\mathrm{p}/R_1$ ratio (Figure \ref{pic:l2_norm_scatterplot}) for the two hydro models Hydro I and Hydro II. We can observe an exponential dependence for the model Hydro I, in which the accuracy of \texttt{FLAIM} with respect to an equivalent \texttt{B2} simulation rapidly worsens for increasing values of the $\Delta_\mathrm{p}/R_1$ ratio, i.e. for relatively thick pushers. Conversely, for the model Hydro II the correlation between the $L_2$ norm and the ratio $\Delta_\mathrm{p}/R_1$ is nearly disappeared. The residual error cannot be clearly ascribed to a specific aspect of the modelling, but it is most likely a combination of the intrinsic differences between \texttt{FLAIM} and \texttt{B2} (e.g., bulk vs. local thermal equilibration, type of discretisation, numerical schemes).

\subsection{Qualitative comparison: temporal profiles}
\noindent Whilst the analysis of the $L_2$ norms can give us a quantitative idea of the average distance between the profiles predicted by \texttt{FLAIM} and \texttt{B2}, the difference is well-demonstrated by some selected simulations. We have chosen three points out of the original parameter scan:

\begin{itemize}
	\item Case 1: $R_1=450~\unit{\mu m}; \Delta_\mathrm{p}=30~\unit{\mu m}; v=200~\unit{km/s}$;
	\item Case 2: $R_1=300~\unit{\mu m}; \Delta_\mathrm{p}=65~\unit{\mu m}; v=200~\unit{km/s}$;
	\item Case 3: $R_1=150~\unit{\mu m}; \Delta_\mathrm{p}=100~\unit{\mu m}; v=200~\unit{km/s}$.
\end{itemize}

\begin{figure}
	\centering
	{\includegraphics[width=1\textwidth,height=0.35\textheight]{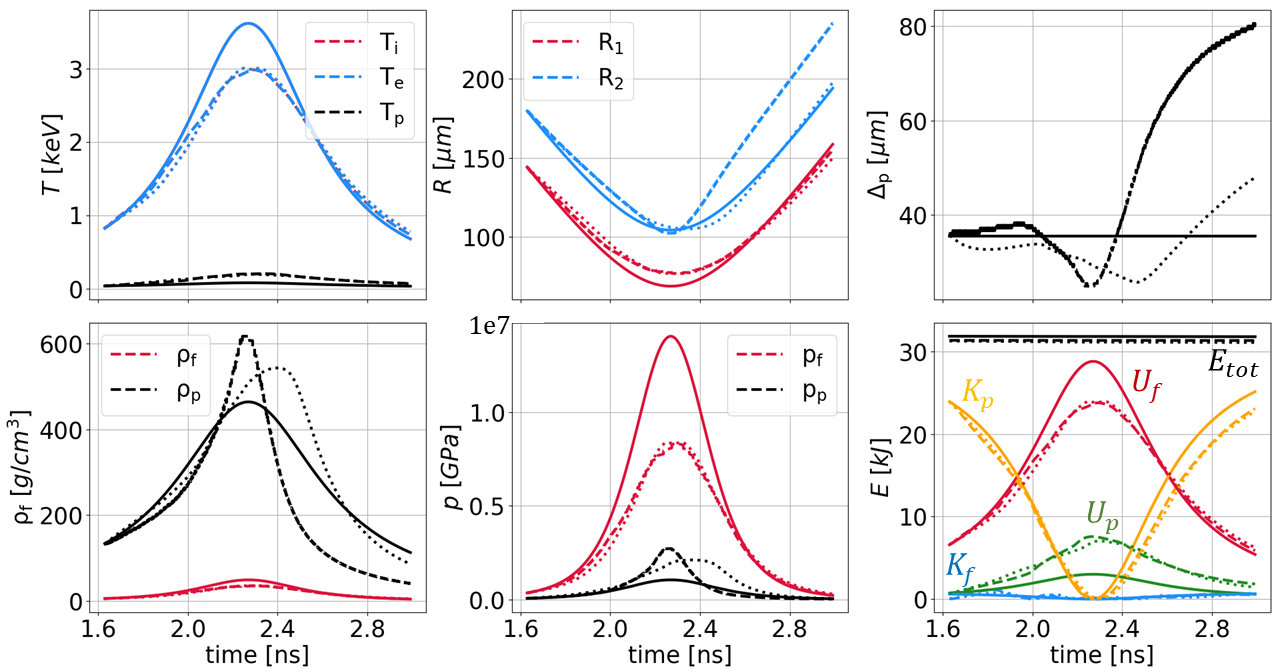}}
	\caption{Temporal profiles of temperatures $T_\mathrm{i}, T_\mathrm{e}, T_\mathrm{p}$, radii $R_1, R_2$, pusher thickness $\Delta_\mathrm{p}$ (top row). Fuel and pusher densities $\rho_\mathrm{f}, \rho_\mathrm{p}$, pressures $p_\mathrm{f}, p_\mathrm{p}$, and energy contributions (kinetic $K_i$, internal $U_i$, total $E_{\mathrm{tot}} = \sum_{i}^{}U_i+K_i$, for $i \in \{\mathrm{f, p}\}$) (bottom row). Solid line is \texttt{FLAIM} with model Hydro I, dotted line is \texttt{FLAIM} with model Hydro II, dashed line is \texttt{B2}. Case 1.}
	\label{pic:flaim_vs_b2_good}
\end{figure}

\begin{figure}
	\centering
	{\includegraphics[width=1\textwidth,height=0.35\textheight]{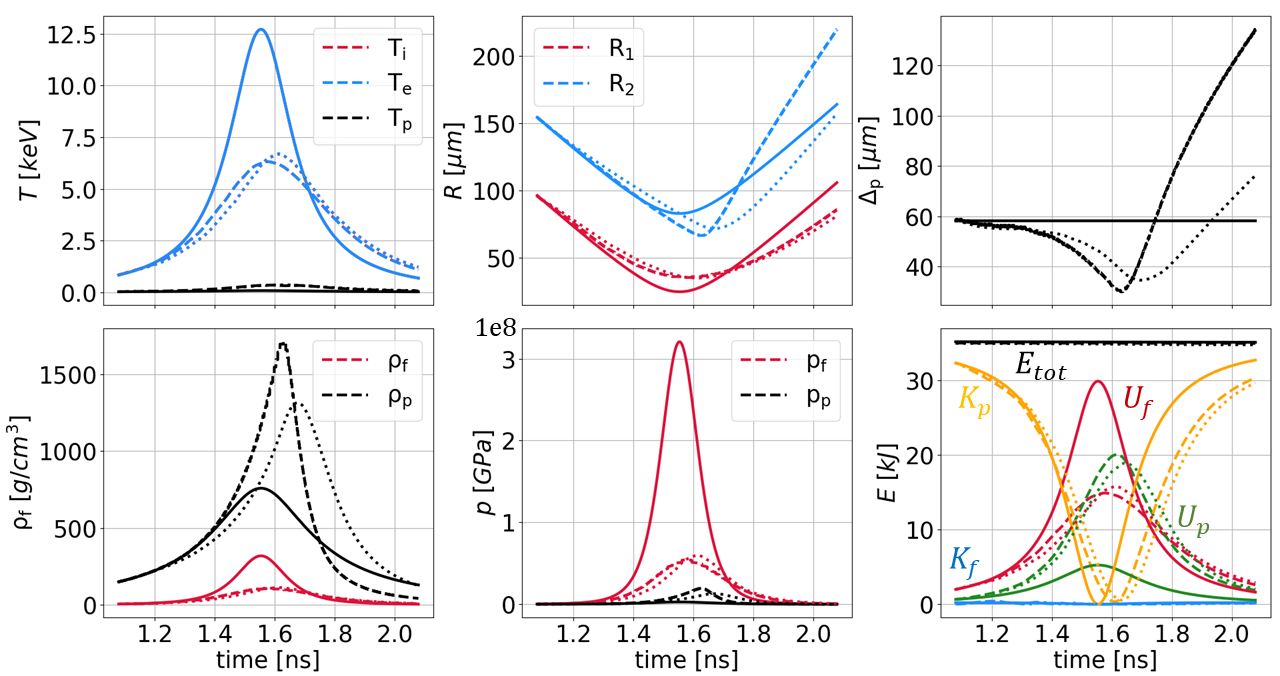}}
	\caption{Temporal profiles of temperatures $T_\mathrm{i}, T_\mathrm{e}, T_\mathrm{p}$, radii $R_1, R_2$, pusher thickness $\Delta_\mathrm{p}$ (top row). Fuel and pusher densities $\rho_\mathrm{f}, \rho_\mathrm{p}$, pressures $p_\mathrm{f}, p_\mathrm{p}$, and energy contributions (kinetic $K_i$, internal $U_i$, total $E_{\mathrm{tot}} = \sum_{i}^{}U_i+K_i$, for $i \in \{\mathrm{f, p}\}$) (bottom row). Solid line is \texttt{FLAIM} with model Hydro I, dotted line is \texttt{FLAIM} with model Hydro II, dashed line is \texttt{B2}. Case 2.}
	\label{pic:flaim_vs_b2_avg}
\end{figure}

\begin{figure}
	\centering
	{\includegraphics[width=1\textwidth,height=0.35\textheight]{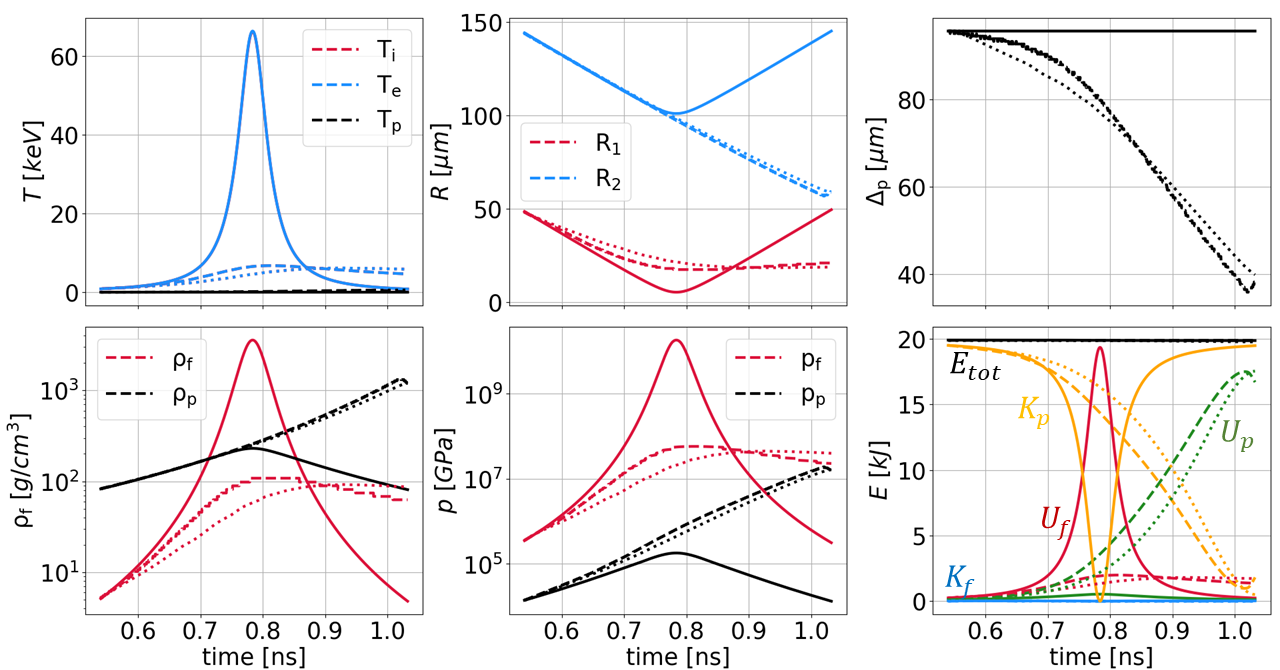}}
	\caption{Temporal profiles of temperatures $T_\mathrm{i}, T_\mathrm{e}, T_\mathrm{p}$, radii $R_1, R_2$, pusher thickness $\Delta_\mathrm{p}$ (top row). Fuel and pusher densities $\rho_\mathrm{f}, \rho_\mathrm{p}$, pressures $p_\mathrm{f}, p_\mathrm{p}$, and energy contributions (kinetic $K_i$, internal $U_i$, total $E_{\mathrm{tot}} = \sum_{i}^{}U_i+K_i$, for $i \in \{\mathrm{f, p}\}$) (bottom row). Solid line is \texttt{FLAIM} with model Hydro I, dotted line is \texttt{FLAIM} with model Hydro II, dashed line is \texttt{B2}. Case 3.}
	\label{pic:flaim_vs_b2_bad}
\end{figure}

\noindent These cases are shown in Figures \ref{pic:flaim_vs_b2_good}, \ref{pic:flaim_vs_b2_avg}, \ref{pic:flaim_vs_b2_bad}, comparing the temporal profiles of key physical quantities for \texttt{B2} (dashed line) vs. \texttt{FLAIM} using the model Hydro I (solid line) and Hydro II (dotted line).  The most evident aspect to notice (for all cases) is that the pusher thickness predicted by \texttt{B2} strongly deviates from the constant profile assumption underlying the Hydro I model, particularly for thick pushers: the shell actually undergoes a significant compression during the implosion phase, followed by a strong expansion after stagnation. Elucidation of the origin of this behaviour is provided by examination of the evolution of the energetics in Figure \ref{pic:flaim_vs_b2_bad}. A clear correlation between the internal energy of the fuel and kinetic energy of the pusher can be seen for the Hydro I model, indicating that the latter is almost entirely used to compress the capsule. The same profiles predicted by \texttt{B2} are qualitatively different: in this case, a significant fraction of the pusher kinetic energy is consumed to compress the pusher \textit{itself}, leaving only a minor contribution for the fuel compression. This explains why Hydro I tends to grossly over-predict the $pdV$ work on the fuel and this is visible from the fuel radius $R_1$ profile, as well as the fuel density $\rho_\mathrm{f}$, pressure $p_\mathrm{f}$ and temperature $T_\mathrm{f}$ in Figures \ref{pic:flaim_vs_b2_avg}, \ref{pic:flaim_vs_b2_bad}. Forcing the imploding shell to maintain a constant thickness puts a severe limit to its compressibility, leaving a kinetic energy reservoir that is inevitably spent on the capsule compression. This phenomenon is less evident for thin shells, but has a strong impact for thicker pushers (in Figure \ref{pic:flaim_vs_b2_bad}, $\Delta^{\mathrm{min}}_\mathrm{p}/\Delta^0_\mathrm{p}\approx0.5$) leading to the conclusion that, whilst the assumption of constant thickness can be reasonable for thin pushers, it rapidly breaks down for thicker ones. This is the same issue encountered for the very simplistic model Hydro 0 (Table \ref{table:hydro_models}), and it demonstrates that even an improved model adopting a constant-thickness compressible pusher is still far from accurate.  A similar conclusion has been reached by previous works \cite{betti2001hot, betti2002deceleration} in the context of hot-spot ignition regarding the inadequacy of incompressible thin shell models (equivalent to the model Hydro 0). The model Hydro I describes an intermediate situation (a pusher with \textit{limited} compressibility), but the physical interpretation is analogous.

Analysing the results from Hydro II, we can see that the fuel compression is described much more accurately. The profiles show good agreement with respect to \texttt{B2}. The pusher thickness profiles predicted by \texttt{FLAIM} follow the \texttt{B2} ones very closely, exhibiting a major compression during the implosion, particularly for thick pushers. The shell compression occurs at the expense of its own kinetic energy, which leaves less margin to compress the fuel.  The energy distribution between fuel and pusher is also correctly predicted. \\ An alternative way to interpret these results focuses on the shock propagation in the pusher. Whilst for thin shells the information is rapidly propagated between the inner and the outer surface, thick shells require more time for this to occur. This leads to a \textit{reduced} compressive work on the fuel provided only by the shocked region of the pusher. Moreover, thick shells are also characterised by a strong pressure gradient \cite{Molvig_PhysPlasmas_2018} (directed inwards), which further contributes to the pusher deceleration. Differently from the Hydro I model, the Hydro II model naturally describes the radial pressure profile and is capable of tracking shocks in the materials (Section \ref{section:mathematical_model}), capturing this dynamics correctly.

\section{Comparison with 1D \replaced[id=P2]{fully-integrated physics}{full-physics} simulations}
\label{sec:multi_physics_simulations}
\noindent In the previous section we analysed in detail the level of accuracy of the hydrodynamics implemented in \texttt{FLAIM}, finding that the approximation of a flat velocity profile in the pusher (i.e., constant thickness) had a detrimental impact on the fidelity of the compression physics. In this section, we present a comparison with \texttt{B2} for burning fuel capsules, to investigate the actual impact that an incorrect treatment of the compression dynamics has on the ignition and burn physics of the target. The scan defined in Table \ref{table:b2_scan} \added[id=P2]{is defined as a fully-integrated physics scan, i.e., } run activating the burn operator and the heat losses (conduction and radiation) for both \texttt{FLAIM} and \texttt{B2}. The electron-ion split-factor is based on the Fraley's model \cite{Fraley_PhysFluids_1974} for both codes. The $\alpha$ particle escape fraction in \texttt{FLAIM} is based on the model of Khrokhin and Rozanov \cite{krokhin1973escape}, whilst in \texttt{B2} it is naturally accounted for  solving a diffusion equation for the $\alpha$ particle transport \cite{atzeni1981diffusive}.

Instead of relying on $L_2$ norms, the comparison in this case is carried out in terms of heatmaps of key burn properties of interest to determine the target performance, namely the final neutron yield $E_\mathrm{n}$ and the deuterium burn fraction $\Phi_\mathrm{D}$, respectively defined as

\begin{equation}
	E_\mathrm{n}=\sum_{i}^{N_\mathrm{R}}\mathcal{R}_i K_{\mathrm{n}, i} \\,
	\label{neutron_energy}
\end{equation} 

\begin{equation}
	\Phi_\mathrm{D}=1-\frac{n_\mathrm{D}}{n_\mathrm{D}^0} \,.
	\label{burn_fraction}
\end{equation} 

\begin{figure}
	\centering
	{\includegraphics[width=1\textwidth,height=0.4\textheight]{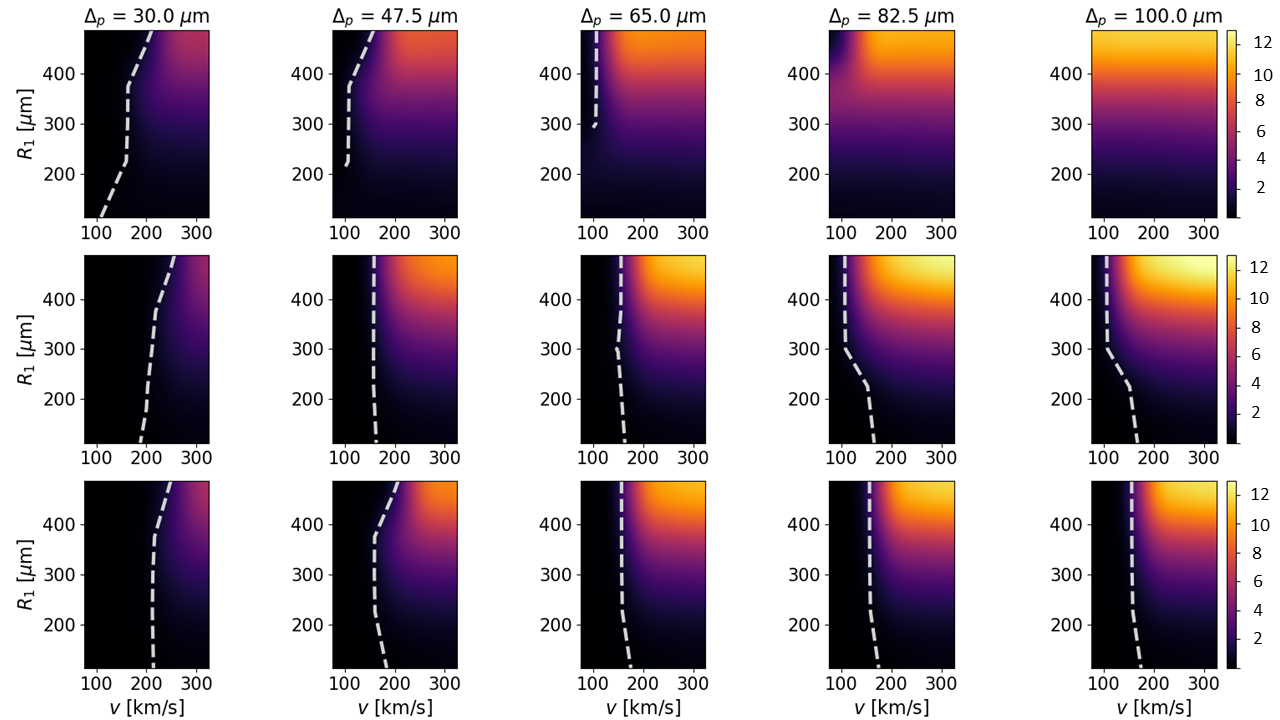}}
	\caption{Heatmaps of final neutron yield $E_\mathrm{n}$ [$\unit{MJ}$] for the \deleted[id=P2]{full-physics} simulation scan defined in Table \ref{table:b2_scan}.  Hydro I  \texttt{FLAIM} (top row), Hydro II \texttt{FLAIM} (middle row) and \texttt{B2} (bottom row).  The dashed lines indicate the iso-contour of constant burn fraction $\Phi_\mathrm{D}=0.05$.}
	\label{pic:neutron_yield_heat_map}
\end{figure}

\begin{figure}
	\centering
	{\includegraphics[width=1\textwidth,height=0.4\textheight]{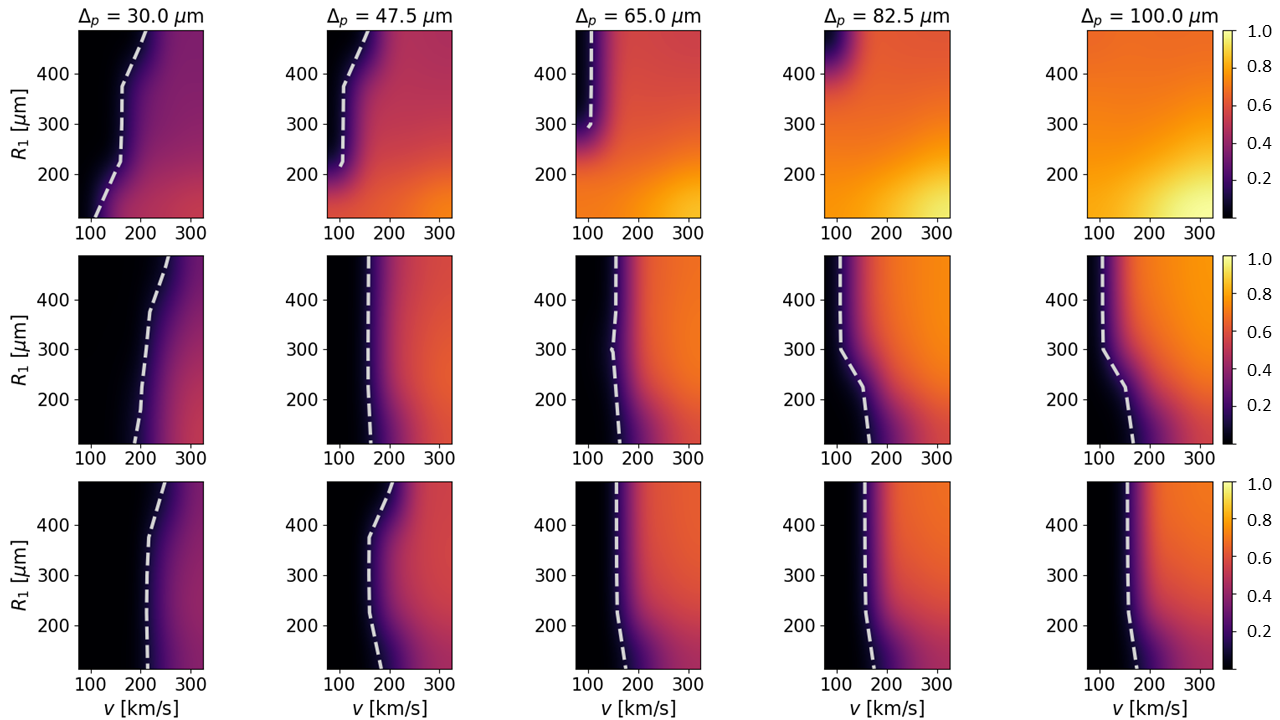}}
	\caption{Heatmaps of final deuterium burn fraction $\Phi_\mathrm{D}$ for the \deleted[id=P2]{full-physics} simulation scan defined in Table \ref{table:b2_scan}.  Hydro I  \texttt{FLAIM} (top row), Hydro II \texttt{FLAIM} (middle row) and \texttt{B2} (bottom row).  The dashed lines indicate the iso-contour of constant burn fraction $\Phi_\mathrm{D}=0.05$.}
	\label{pic:burn_fraction_heat_map}
\end{figure}

\noindent Figures \ref{pic:neutron_yield_heat_map} and \ref{pic:burn_fraction_heat_map} report the surfaces of $E_\mathrm{n}$ and $\Phi_\mathrm{D}$ in the parameter space predicted by \texttt{FLAIM} with Hydro I  (top row), \texttt{FLAIM} Hydro II (middle row) and \texttt{B2} (bottom row).

The \texttt{B2} maps show a distinct separation between a non-burning region and a strongly burning region, in which the separatrix (white dashed line) identifies the iso-contour corresponding to $\Phi_\mathrm{D}=0.05$. This line can be interpreted as the minimum implosion velocity $v^*$ (function of $R_1, \Delta_\mathrm{p}$)  required to ignite and burn a significant fraction of the fuel. The choice of using a $\Phi_\mathrm{D}$ iso-contour was motivated by the fact that we found the volume ignition criterion by Molvig et al. \cite{Molvig_PhysRevLett_2016} (Equation \ref{molvig_criterion}) to be insufficient for defining sensible separatrices in the maps. In particular, Equation \ref{molvig_criterion} labels as igniting \textit{all} the simulations of the scan (Table \ref{table:b2_scan}). While we recognise the intention behind this criterion is to identify the moment of upstream ignition for a sans-losses implosion (like Revolver), it ultimately fell short in effectively distinguishing between igniting and non-igniting regions within the parameter space. Hence, the choice to adopt a simple $\Phi_\mathrm{D}=0.05$ iso-contour, representing $\sim1/10$ of the fuel burn fraction of Revolver \cite{Molvig_PhysPlasmas_2018}. Given the steep gradient of the surface $\Phi_\mathrm{D}\left(R_1, v, \Delta_\mathrm{p}\right)$ (characteristic of ignition phenomena), no noticeable change in the separatrix position was observed when varying the iso-contour value within the range $\{0.01-0.15\}$.

Looking at the \texttt{B2} maps (bottom row), for thin shells, $v^*$ tends to decrease with the pusher thickness (from $\sim 210~\unit{km/s}$ for a $30~\unit{\mu m}$ pusher to $\sim 180~\unit{km/s}$ for a $65~\unit{\mu m}$ pusher), which can be explained by the higher kinetic energy available to compress the fuel. However, a plateau is visible for $\Delta_\mathrm{p} \gtrsim 65~\unit{\mu m}$: further thickening the pusher after this point does not change the ignition requirements in terms of implosion velocity. This means that the target is able to convert only a fraction of the pusher kinetic energy to fuel internal energy and the rest is simply taken up by the shell self-compression. In the same region ($\Delta_\mathrm{p} \gtrsim 65~\unit{\mu m}$), whilst the neutron yield (Figure \ref{pic:neutron_yield_heat_map}) increases with the capsule size (because more fuel is available), the final burn fraction (Figure \ref{pic:burn_fraction_heat_map}) is roughly independent of it.

\texttt{FLAIM} run with Hydro II (middle row) generally compares well with \texttt{B2}: the lines representing $v^*$ are in good agreement and clearly separate two similar regions of the parameter space. For thicker pushers, \texttt{FLAIM} exhibits a more pronounced dependence on $R_1$, whereas $\texttt{B2}$ provides a quasi-vertical separatrix. As reported in Section \ref{subsec:heat_losses}, this is attributable to the simplified thermal conduction model, that tends to overestimate the heat losses at small fuel radii (Equation \ref{thermal_conduction_form}). Both the neutron yield and the burn fraction are in good agreement, even though slightly overestimated, especially at high implosion velocity. Finally, \texttt{FLAIM} tends to underevaluate $v^*$ for thicker pushers, showing a premature ignition with respect to \texttt{B2}. Overall, the comparison is satisfactory, showing that \texttt{FLAIM} can predict with sufficient accuracy the ignition surfaces in the parameter space when adopting the model Hydro II for the hydrodynamics.

Conversely, the heatmaps predicted using  the model Hydro I (top row) are qualitatively different. The comparison is acceptable for very thin pushers, but rapidly worsens for thicker ones. In particular, for $\Delta_\mathrm{p}\geq82~\unit{\mu m}$ the strongly-burning region covers almost the whole parameter space, with unrealistic burn fractions up to $\Phi_\mathrm{D}\approx 1$ predicted. This behaviour is a consequence of the incorrect description of the hydrodynamics resulting from assuming a constant thickness shell, that we extensively analysed in Section \ref{section:comparison_b2}. The overestimation of the fuel compression has a direct impact on the burn physics due to its quadratic dependency on the mass density, leading to the misleading interpretation that thick pushers are more efficient in compressing the capsule because of their large kinetic energy. In reality, only a fraction of this kinetic energy can be effectively coupled to the the fuel, whilst the rest is actually spent for shell self-compression. Interestingly, the \texttt{B2} maps show that this fraction is independent of the pusher thickness, and the \texttt{FLAIM}  scan with Hydro II captures this behaviour. The tendency of the model Hydro I \deleted[id=P2]{, when included in a full-physics \texttt{FLAIM} simulation,} to predict ignition and strong burn in regions where higher-fidelity simulations fail to do so is insidious and the worst possible scenario for use in optimisations because it is fundamentally \textit{optimistic}. Using a reduced  model for large scans, one naturally expects some discrepancy in predicting ignition regions, but one would rather have a conservative version of it, in which ignition and burn are not predicted where they actually should be. This allows for the  design and optimisation of experiments with some margin for the expected performance, even accounting for the inevitable uncertainties of a simplified modelling approach.

\subsection{Analysis on the role of heat losses}
\label{subsec:heat_losses}

\begin{table}
	\centering
	\begin{tabular}{lll}
		\toprule
		& \quad \quad \texttt{B2} scan & n points \\
		\midrule
		$R_1$ $[\unit{\mu m}]$ & \quad \quad $150-450$ & \quad \quad 10\\ 
		\midrule
		$\Delta_\mathrm{p}$ $[\unit{\mu m}]$ & \quad \quad $82.5-100$ & \quad \quad 3\\ 
		\midrule
		$v$ $[\unit{km/s}]$ & \quad \quad $100-300$ & \quad \quad 10\\ 
		\bottomrule
	\end{tabular}
	\caption{High-resolution \texttt{B2} simulation scan (300 simulations, for each coordinate the n points are equidistant) in terms of initial fuel radius $R_1$, initial pusher thickness $\Delta_\mathrm{p}$ and initial implosion velocity $v=v_1=v_2$, defining a sub-region of the space in Table \ref{table:b2_scan}.}
	\label{table:b2_scan_higher_res}
\end{table}

\noindent Although we have demonstrated that the efficacy of \texttt{FLAIM} as a predictive surrogate  using the model Hydro II, some discrepancies persist across the design space. Firstly, whilst the accuracy is satisfactory for small thickness pushers (and much more improved for thick ones), it is possible to observe a region ($120 ~ \unit{km/s} \lesssim v^* \lesssim 170 ~ \unit{km/s}$) where \texttt{FLAIM} still incorrectly predicts ignition and burn. Secondly, even when both models roughly agree on the shape of the ignition region, \texttt{FLAIM} tends to slightly overpredict both neutron yield and burn fraction for thick pushers. To complete the analysis, it is worth investigating the reason behind these inconsistencies turning our attention to the heat loss mechanisms. As reported in Section \ref{section:mathematical_model}, thermal conduction is implemented in a simplified manner (Equation \ref{thermal_conduction_form}), as well as the radiation loss (Equation \ref{radiation_loss}). Moreover, the wall  (which ultimately governs the temperature difference that drives the heat fluxes) is also modelled with a  simplified approach to avoid the solution of diffusion equation (Equation \ref{radiation_diffusion}). Heat losses play a key role in determining the ignition of the fuel capsule and, whilst their accurate modelling is (deliberately) out of the scope of a work on a simplified ignition model like \texttt{FLAIM}, it is valuable to analyse the impact that a simplified treatment has on the model accuracy.

\begin{figure}
	\centering
	{\includegraphics[width=0.85\textwidth,height=0.42\textheight]{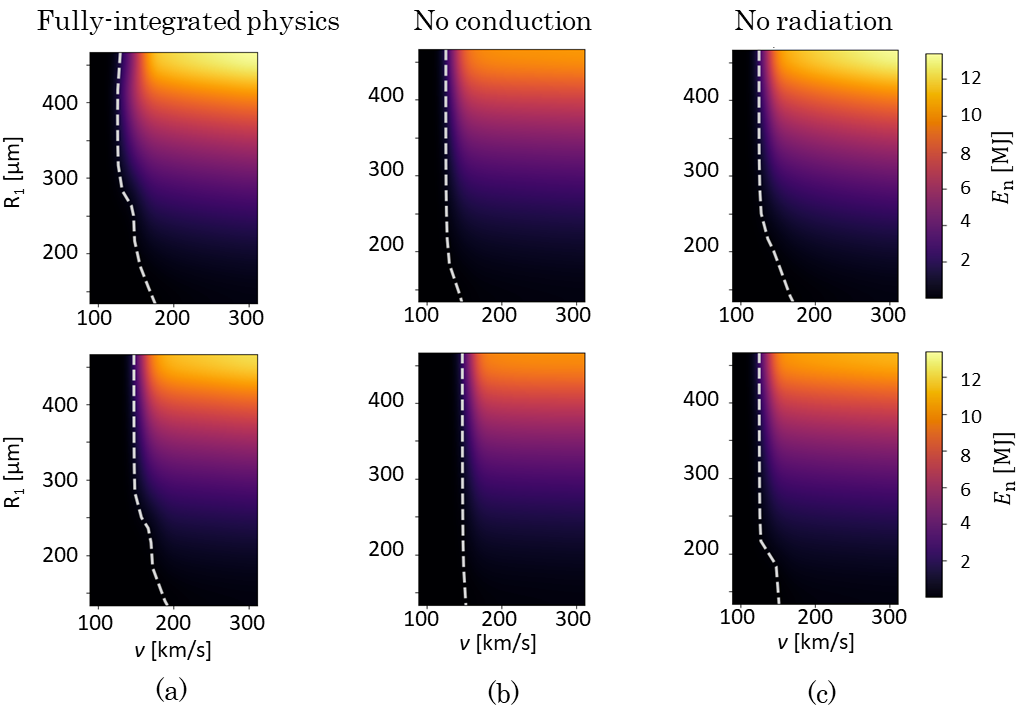}}
	\caption{Heatmaps of neutron energy $E_\mathrm{n}$ for $\Delta_\mathrm{p}=82.5~\unit{\mu m}$ for  \replaced[id=P2]{fully-integrated physics}{full physics} simulations (a), \deleted[id=P2]{full-physics simulations}without thermal conduction (b) and \deleted[id=P2]{full-physics simulations} without radiation loss (c).  \texttt{FLAIM} results (top row) vs. \texttt{B2} results (bottom row). The dashed lines indicate the iso-contour of constant burn fraction $\Phi_\mathrm{D}=0.05$.}
	\label{pic:heat_losses_picture_825um}
\end{figure}

To this purpose, we ran a \texttt{B2} scan removing each heat loss mechanism independently to investigate their impact in isolation:

\begin{enumerate}
	\item The first scan is run removing radiation loss, leaving only thermal conduction as the sole heat loss mechanism;
	\item The second scan is run removing thermal conduction (both ion and electron), leaving only radiation loss as the sole heat loss mechanism.
\end{enumerate}

\noindent Since thick pushers are most relevant for the discussion, we ran a scan in the sub-region $82.5~\unit{\mu m} \leq \Delta_\mathrm{p} \leq 100 ~ \unit{\mu m}$, increasing the resolution in terms of fuel radius $R_1$ and implosion velocity $v$ (from 5 to 10 points for each coordinate). The scan now includes 300 simulations in a sub-region of the parameter space defined in Table \ref{table:b2_scan} and is summarised in Table \ref{table:b2_scan_higher_res}. This scan is run for the three cases: (i) \replaced[id=P2]{with all physical operators activated}{full-physics scan} (i.e., hydrodynamics, burn, conduction, radiation)\footnote{This scan was shown in Section \ref{section:comparison_b2}, but it is run again at the same resolution defined in Table \ref{table:b2_scan_higher_res}.}; (ii) \deleted[id=P2]{full-physics scan} without thermal conduction (i.e., hydrodynamics, burn, radiation); (iii) \deleted[id=P2]{full-physics scan} without radiation loss (i.e., hydrodynamics, burn, conduction). Thermal equilibration and wall operators are always included. The hydrodynamic model is set to Hydro II (Table \ref{table:hydro_models}).

\begin{figure}
	\centering
	\subfloat[]
	{\includegraphics[width=.49\textwidth,height=0.23\textheight]{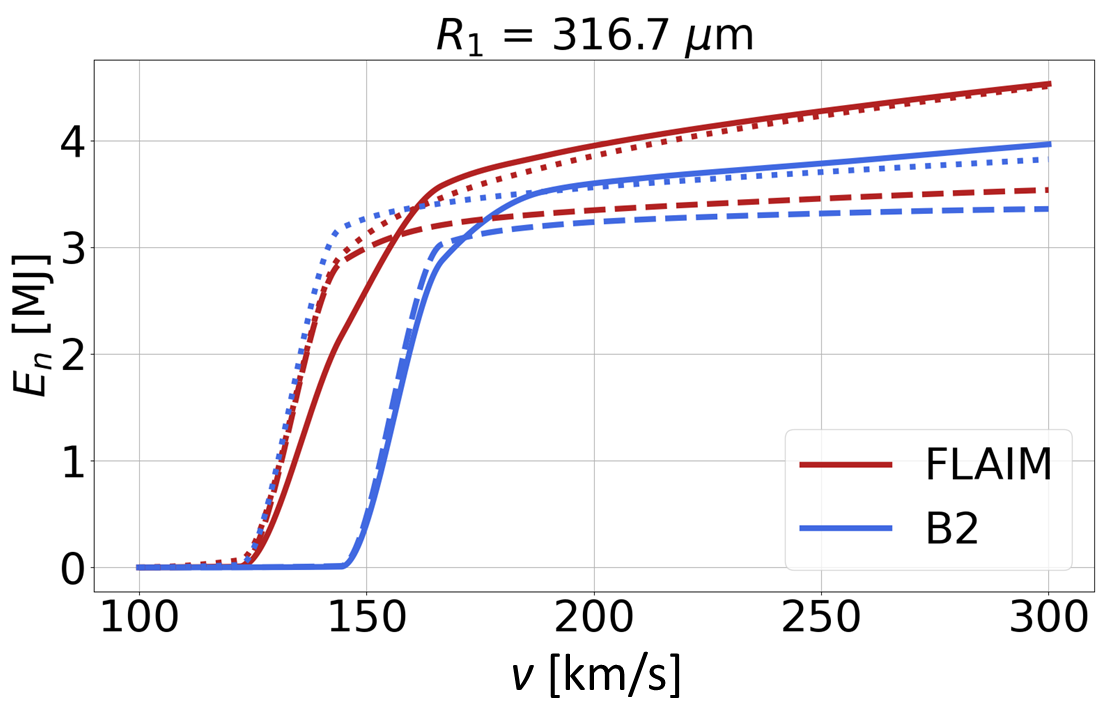}} ~
	\subfloat[]
	{\includegraphics[width=.49\textwidth,height=0.23\textheight]{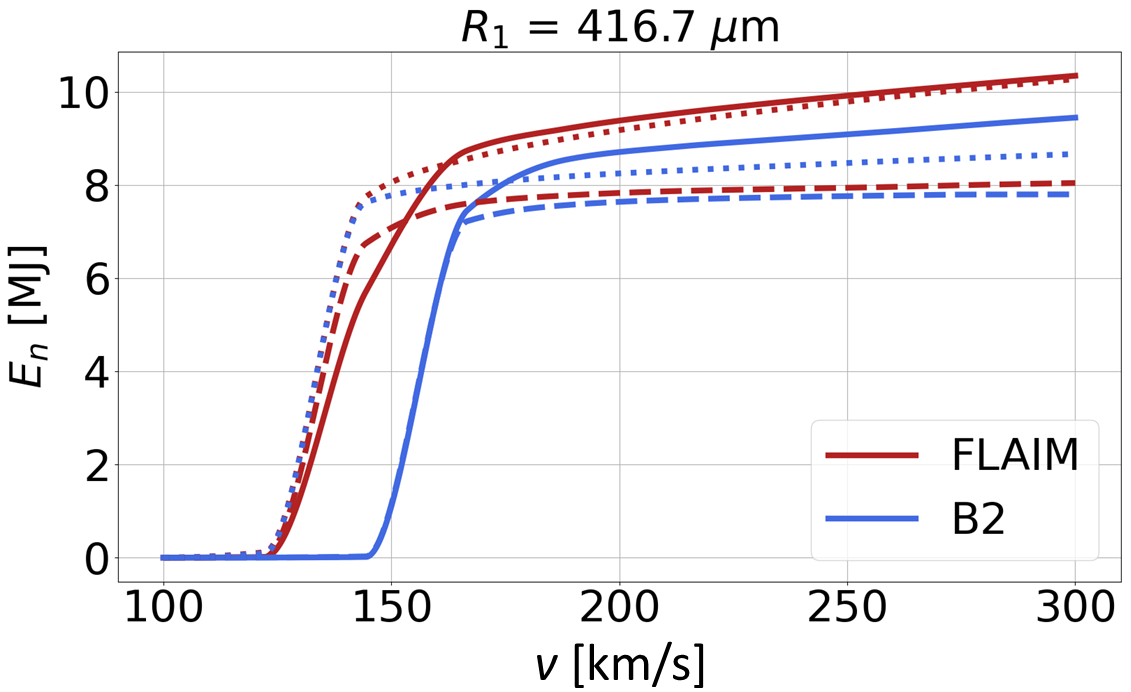}}
	\caption{Profiles of neutron energy $E_\mathrm{n}$ along iso-lines of constant fuel radius $R_1=316.7 ~\unit{\mu m}$ (a) and $R_1=416.7 ~\unit{\mu m}$ (b) for the heatmap at $\Delta_\mathrm{p}=82.5~\unit{\mu m}$ (Figure \ref{pic:heat_losses_picture_825um}). Full-physics (solid line), no-conduction (dashed line) and no-radiation (dotted line).}
	\label{pic:slices}
\end{figure}

Figures \ref{pic:heat_losses_picture_825um} a, b, c report the neutron yield $E_\mathrm{n}$ maps for the three cases respectively, for \texttt{FLAIM} and \texttt{B2} at $\Delta_\mathrm{p}=82.5~\unit{\mu m}$. Both codes predict a lower neutron yield when heat losses are removed, due to the higher  adiabat and the reduced compressibility of the fuel capsule.  As previously discussed, the \deleted[id=P2]{full-physics} \texttt{FLAIM} scan presents two aspects of discrepancy with respect to \texttt{B2}. Firstly, a region of premature ignition ($120 ~ \unit{km/s} \lesssim v^* \lesssim 170 ~ \unit{km/s}$) can be observed. At higher resolution, its width is slightly reduced   ($120 ~ \unit{km/s} \lesssim v^* \lesssim 160 ~ \unit{km/s}$), but still clearly visible. Secondly, a general overestimation of $E_\mathrm{n}$ for large capsule radii is noticeable.   Analysing the map run without thermal conduction (b), one immediately notices that the agreement in terms of $E_\mathrm{n}$ between \texttt{FLAIM} and \texttt{B2} is now excellent for $v^* \gtrsim 160 ~ \unit{km/s}$. The dependency of $v^*$ on $R_1$ visible in the \deleted[id=P2]{full-physics} maps \added[id=P2]{in Figure \ref{pic:heat_losses_picture_825um}a} has almost disappeared, providing a quasi-vertical separatrix for both codes.  However, the spurious burning region ($120 ~ \unit{km/s} \lesssim v^* \lesssim 160 ~ \unit{km/s}$) that was visible \replaced[id=P2]{for the \texttt{FLAIM} scan in Figure \ref{pic:heat_losses_picture_825um}a}{in the \texttt{FLAIM} full-physics maps} is still present. This leads to the conclusion that thermal conduction dominates the losses in the burning regime since most of the impact of its removal is localised in this region of the parameter space, affecting in a minor way the position of the separatrix.

This behaviour is confirmed analysing the scan in which radiation is removed from the system (c): in this case we observe the opposite scenario, in which the general overprediction of $E_\mathrm{n}$ in the burning regime remains unchanged, whilst $v^*$ is now in good agreement to what predicted by \texttt{B2} (with minor deviations at small radii due to the conduction model). This indicates that radiation loss is the main heat loss mechanism during the implosion phase (defining whether it would lead to ignition or not) and its removal allows us to accurately predict the location of the strong-burning region with respect to \texttt{B2}.

\begin{figure}
	\centering
	\subfloat[]
	{\includegraphics[width=.47\textwidth,height=0.19\textheight]{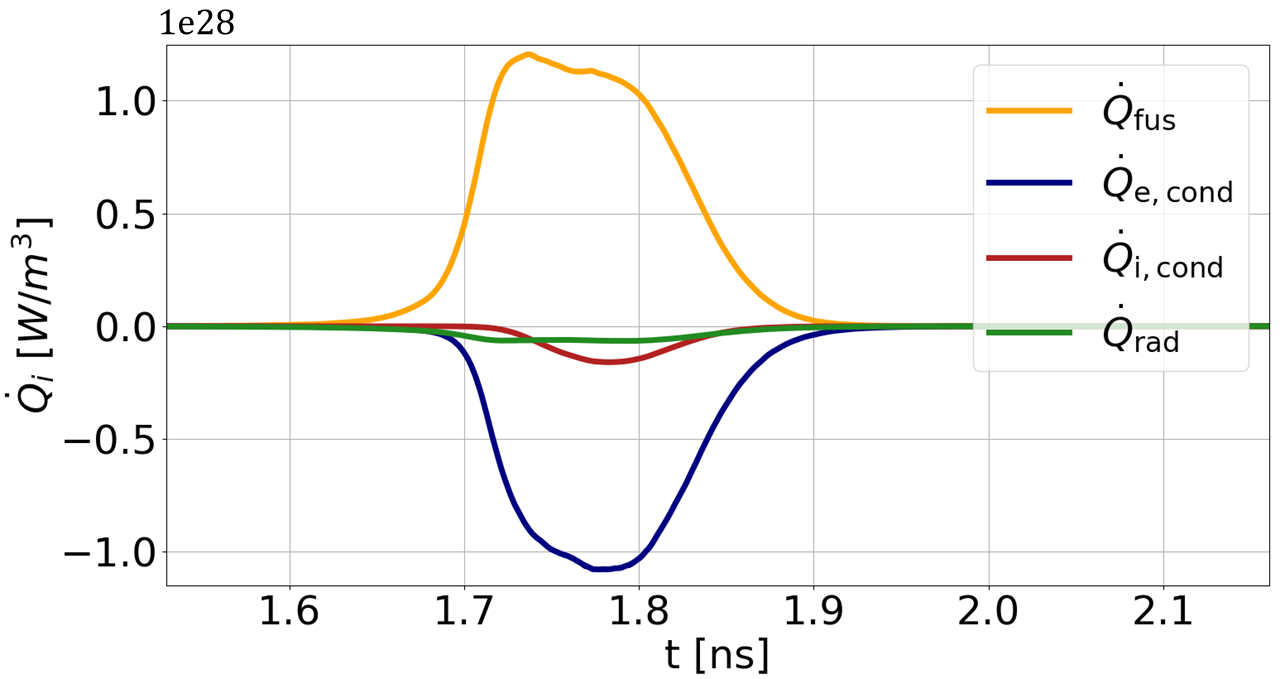}} ~
	\subfloat[]
	{\includegraphics[width=.47\textwidth,height=0.19\textheight]{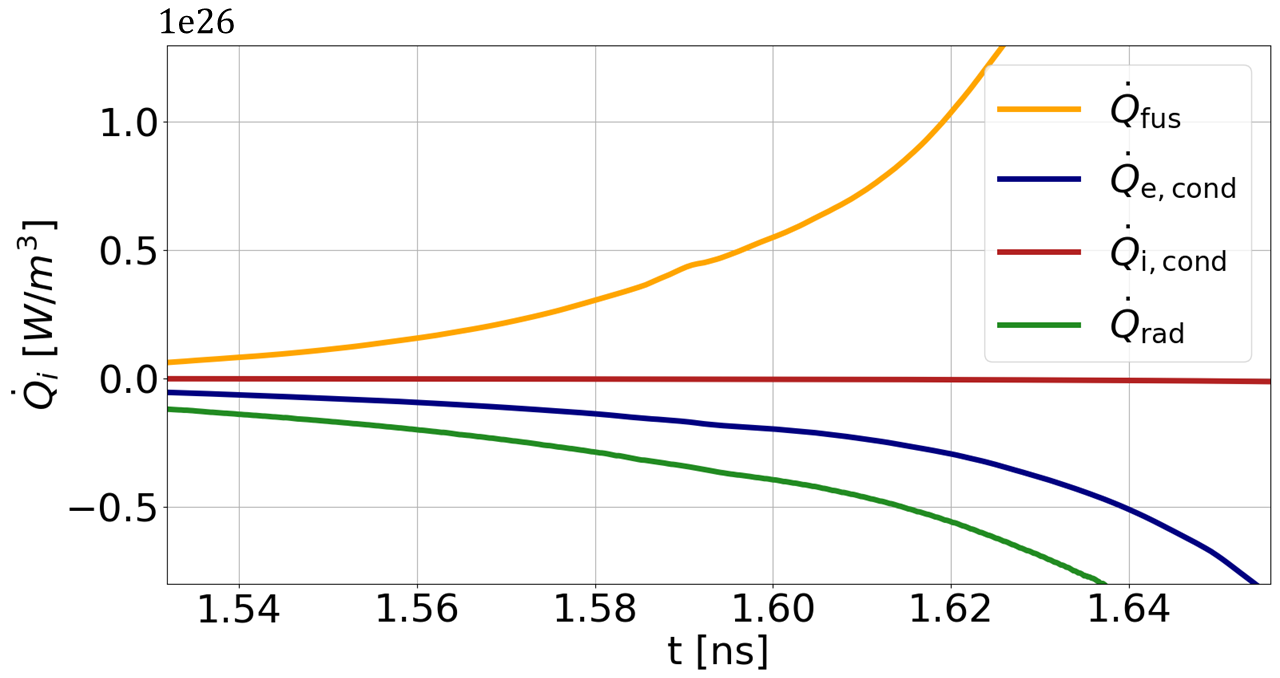}}
	\caption{Plots of power density contributions (burn $\dot{Q}_{\mathrm{fus}}$, electron conduction $\dot{Q}_{\mathrm{e, cond}}$, ion conduction $\dot{Q}_{\mathrm{i, cond}}$ and radiation $\dot{Q}_{\mathrm{rad}}$) to the fuel internal energy density $u_f$ for the Revolver design \cite{Molvig_PhysRevLett_2016} simulated with \texttt{FLAIM} using Hydro II. Figure (b) reports a detail of Figure (a) in the range $1.54 \lesssim t\lesssim 1.64$ ns, to highlight the implosion phase.}
	\label{pic:power_plot}
\end{figure}

This can be more clearly observed in Figure \ref{pic:slices}, reporting the profiles of $E_\mathrm{n}$ in the velocity space for two selected radii from  Figure \ref{pic:heat_losses_picture_825um} for the three cases (\replaced[id=P2]{fully-integrated physics}{full-physics}, no conduction, no radiation). The visible gap in the implosion velocity required for ignition is unaffected by the removal of thermal conduction, which however significantly improves the agreement in terms of $E_\mathrm{n}$ at high velocity. Conversely, removing radiation closes the gap in terms of $v*$, leaving the discrepancy of $E_\mathrm{n}$ at high velocity unchanged.

This is further supported by plotting the burn, thermal conduction and radiation loss contributions (in terms of power densities) for the Revolver design  \cite{Molvig_PhysRevLett_2016} using \texttt{FLAIM} (Figure \ref{pic:power_plot}). We can clearly see that radiation dominates over conduction during the implosion phase (before ignition), whilst conduction rapidly takes over in the burning regime (largely dominated by the electron component). We found the flux-limiting on thermal conduction (Equation \ref{flux_limited_conduction}) to have no impact on these results. This tells us that the simplified way in which the heat losses are treated is mainly responsible for the residual discrepancy we observe in the ignition maps. The burn operator in \texttt{FLAIM} shows virtually no difference with respect to the 1D description of the burn implemented in \texttt{B2}, suggesting very uniform  conditions in the fuel capsule after ignition (a desirable feature for volume ignition, which is confirmed by inspecting the \texttt{B2} spatial profiles). This represents a significant advantage when dealing with simplified ignition models, because it removes the necessity of capturing the burn physics and the energy deposition in detail, for example, implementing methods for Charged Particle Transport (CPT) \cite{atzeni1981diffusive}.

To conclude, this work also demonstrates that the potentially large discrepancy between a reduced simple model for volume ignition ICF  and a multi-physics high-fidelity code can be removed almost entirely simply by treating the hydrodynamics correctly. As we have shown, heat losses do play an important role, but in the context of reduced models  their accurate description is not strictly necessary since their impact can be roughly captured by simplified methods. Conversely, the same approach is less  advisable when dealing with the hydrodynamic physics,  which has a much stronger impact in determining the fuel conditions necessary for a robust ignition.
\section{Modelling of the Revolver design}
\label{sec:revolver}
\noindent In this last section we use the \texttt{FLAIM} code to model the Revolver  target \cite{Molvig_PhysRevLett_2016, Molvig_PhysPlasmas_2018, Keenan_PhysPlasmas_2020b}, focusing on the last two layers of the structure, i.e. the pusher and the fuel. For a specific target, we assume the required pressure multiplication (originally supplied by the three-shell design) to be provided by FLF's proprietary amplifier technology. The initial conditions for Revolver in terms of geometry, implosion velocity (Table \ref{table:b2_scan}) and material states are reported in Section \ref{subsec:simulation_scan}. The model choices are kept the same as those used in Section \ref{sec:multi_physics_simulations} \replaced[id=P2]{(i.e., with all physics operators activated)}{for the full-physics scan}, except that we consider only the more accurate hydrodynamic model Hydro II (Table \ref{table:hydro_models}). 

\begin{figure}
	\centering
	{\includegraphics[width=1\textwidth,height=0.38\textheight]{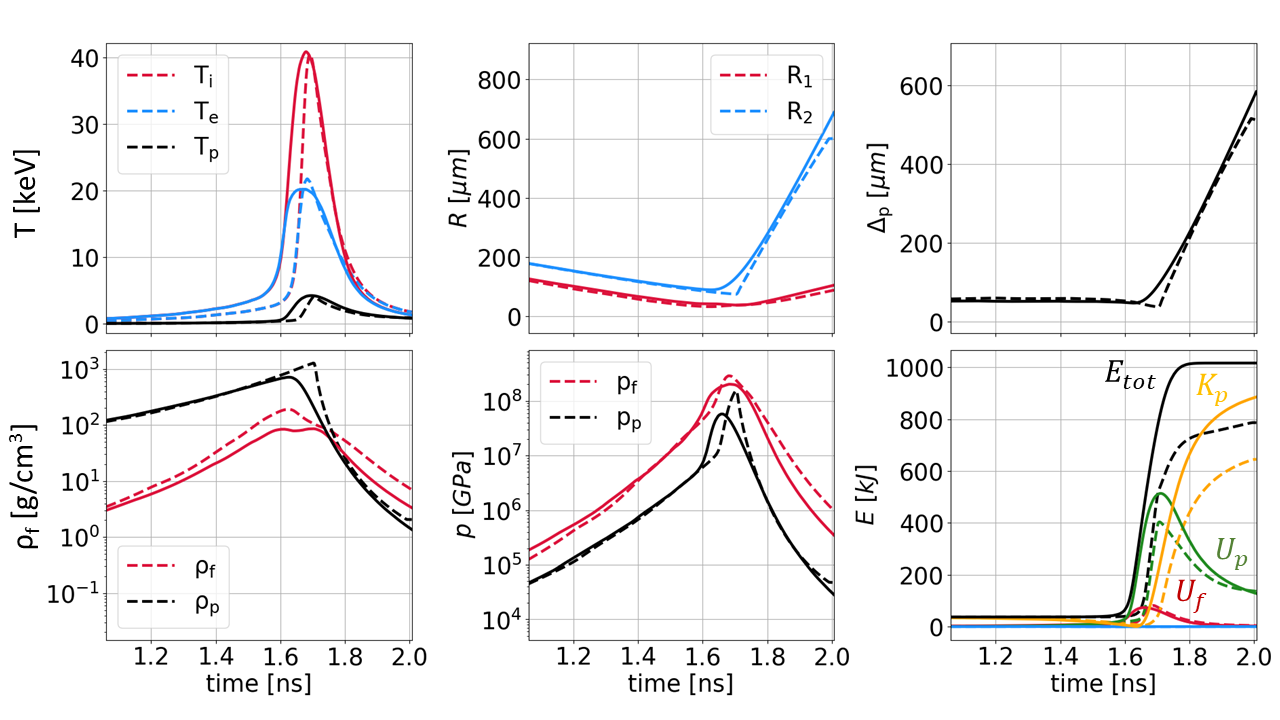}}
	\caption{Temporal profiles of temperatures $T_\mathrm{i}, T_\mathrm{e}, T_\mathrm{p}$, radii $R_1, R_2$, pusher thickness $\Delta_\mathrm{p}$ (top row). Fuel and pusher densities $\rho_\mathrm{f}, \rho_\mathrm{p}$, pressures $p_\mathrm{f}, p_\mathrm{p}$, and energy contributions (kinetic $K_i$, internal $U_i$, total $E_{\mathrm{tot}} = \sum_{i}^{}U_i+K_i$, for $i \in \{\mathrm{f, p}\}$) (bottom row). Solid line is \texttt{FLAIM}, dashed line is \texttt{B2}. Revolver design  \cite{Molvig_PhysPlasmas_2018}.}
	\label{pic:revolver}
\end{figure}

Figure \ref{pic:revolver} reports the temporal profiles of the main physical variables predicted by \texttt{FLAIM} for the two-layer Revolver design, with the \texttt{B2} profiles also reported for reference. The results from \texttt{FLAIM} are compared to the work of Molvig et al. \cite{Molvig_PhysPlasmas_2018}, which is based on 1D \texttt{HYDRA} \cite{marinak2001three} simulations of a laser-driven implosion of the \textit{whole} Revolver target. Following Molvig et al. \cite{Molvig_PhysRevLett_2016}, the ignition criterion is based on the relative contribution of the fusion and the compressive power on the fuel internal energy $u_\mathrm{f}$, due to the negligible radiative losses during the implosion. The earliest time in the simulation where the inequality\footnote{The additional term $\left(1/V_\mathrm{f}\right)\left(\dint V_\mathrm{f}/\dint t\right)u_\mathrm{f}$ in Equation \ref{molvig_criterion} derives from the fact that the hydrodynamic operator evolves $u_\mathrm{f}$ \textit{and} $V_\mathrm{f}$, and both variations must be accounted for.} 

\begin{equation}
	\dot{Q}_{\mathrm{fus}} \geq \frac{1}{4}\left(\dot{Q}_{\mathrm{pdV}} + \frac{1}{V_\mathrm{f}}\diff{V_\mathrm{f}}{t}u_\mathrm{f}\right)
	\label{molvig_criterion}
\end{equation}

\noindent holds defines the ignition time $t_{\mathrm{ign}}$, where $\dot{Q}_{\mathrm{fus}}$ is given by Equation \ref{fuel_internal_energy_burn_equation} and $\dot{Q}_{\mathrm{pdV}}$ by the sum over $\{\mathrm{i, e, r}\}$ and over the $N_{\mathrm{c, f}}$ fuel cells of Equation \ref{energy_equation_lagrangian}. The stagnation time $t_{\mathrm{stagn}}$ is the time at which the interface velocity $v_1=0$, before changing sign as the system expands. Finally, the peak burn time $t_{\mathrm{peak, b}}$ indicates the maximum of $\dot{Q}_{\mathrm{fus}}$. \texttt{FLAIM} includes an  integrated post-processor to extract relevant fiducial times, integrated quantities and general burn metrics. Table \ref{table:revolver} presents the value of some diagnostic variables predicted by \texttt{FLAIM} at $t_{\mathrm{ign}}$, $t_{\mathrm{stagn}}$ and $t_{\mathrm{peak, b}}$, compared with the ones reported in  \cite{Molvig_PhysPlasmas_2018}. If not explicitly reported in the text of the reference work, the relevant quantities are extracted directly from the available plots and are to be considered approximate. Quantities for which it was not immediate to recover a reliable reference value at a specific time are not reported\footnote{Following the work of Molvig et al. \cite{Molvig_PhysPlasmas_2018}, ignition and deceleration onset are regarded as equivalent locations.}.

\begin{table}
	\centering
	\begin{tabular}{lccccccccc}
		\toprule
		& \multicolumn{3}{c}{\quad Ignition} &  \multicolumn{3}{c}{\quad Stagnation} &  \multicolumn{3}{c}{\quad Peak burn} \\
		\cmidrule(l{1cm}r){2-4} \cmidrule(l{1cm}r){5-7} \cmidrule(l{1cm}r){8-10}
		&\quad F & R & $\epsilon$ &\quad F & R & $\epsilon$ &\quad F &  R & $\epsilon$\\
		\midrule
		$t-t_0~~[\unit{ns}]$ & \quad $0.4$ & $0.35$ & $0.14$ &\quad $0.57$ & $0.5$ & $0.14$ &\quad $0.64$ & $0.54$ & $0.18$ \\
		\midrule
		$c_r~[\mathrm{-}]$ & \quad $5.9$ & $6.5$ & $0.09$ &\quad $7.9$ & $9.3$ & $0.15$ &\quad $/$ & $/$ & $/$ \\
		\midrule
		$T_{\mathrm{i}}~[\unit{keV}]$ & \quad $2.9$ & $2.5$ & $0.16$ &\quad $13.2$ & $9.7$ & $0.36$ &\quad $41$ & $35$ & $0.17$ \\
		\midrule
		$T_{\mathrm{e}}~[\unit{keV}]$ & \quad $2.9$ & $2.5$ & $0.16$ &\quad $13.2$ & $9.7$ & $0.36$ &\quad $20$ & $20$ & $0$ \\
		\midrule
		$\rho_{\mathrm{f}}~[\unit{g/cm^3}]$ & \quad $35$ & $40$ & $0.12$ &\quad $84$ & $140$ & $0.4$ &\quad $83$ & $100$ & $0.17$ \\
		\midrule
		$\rho_{\mathrm{p}}~[\unit{g/cm^3}]$ & \quad $427$ & $400$ & $0.07$ &\quad $693$ & $2000$ & $0.6$ &\quad $/$ & $/$ & $/$ \\
		\midrule
		$U_{\mathrm{f}}~[\unit{kJ}]$ & \quad $8$ & $5$ & $0.6$ &\quad $38$ & $28$ & $0.35$ &\quad $69$ & $80$ & $0.14$ \\
		\midrule
		$U_{\mathrm{p}}~[\unit{kJ}]$ & \quad $8$ & $13$ & $0.38$ &\quad $45$ & $36$ & $0.25$ &\quad $/$ & $/$ & $/$ \\
		\midrule
		$K_{\mathrm{p}}~[\unit{kJ}]$ & \quad $21$ & $30$ & $0.3$ &\quad $4$ & $13$ & $0.7$ &\quad $89$ & $80$ & $0.11$ \\
		\bottomrule
	\end{tabular}
	\caption{Results for the Revolver design predicted by  \texttt{FLAIM} (F) and reported in the reference work (R) of Molvig et al. \cite{Molvig_PhysPlasmas_2018}. $c_r=R_1^0/R_1$ is the convergence ratio. The relative error $\epsilon=|y_\mathrm{F}-y_\mathrm{R}|/y_\mathrm{R}$ is also reported. $t_0$ is the time of shock convergence.}
	\label{table:revolver}
\end{table}

From Table \ref{table:revolver} we observe generally a good agreement between \texttt{FLAIM} and the reported number from the reference work. The ignition is a little delayed in \texttt{FLAIM} and occurs at a temperature slightly higher than the comparative value of $\sim 2.5~\unit{keV}$. At ignition, the pusher in \texttt{FLAIM} has $\sim 60\%$ of the initial pusher kinetic energy, as opposed to  $\sim 80\%$ reported by Molvig et al. This suggests a stronger transfer of $K_\mathrm{p}$  to the fuel during the compression phase in \texttt{FLAIM}, and this is confirmed by the small overprediction in the fuel internal energy. Given the minimal impact of conductive losses during this phase (Figure \ref{pic:power_plot}), this probably indicates an underestimation of the radiation losses in \texttt{FLAIM}. This traps more heat inside the capsule and explains the higher ignition temperature and the slightly lower convergence ratio and fuel density (due to the higher adiabat). 

It is worth noticing that the sum of the energies at ignition reported in \cite{Molvig_PhysPlasmas_2018} is more than the initial $K_\mathrm{p}\approx 37~\unit{kJ}$ characteristic of Revolver. This indicates some form of pre-heat during the implosion (e.g., radiation transport from the drive on the outer shells) that is not captured in \texttt{FLAIM} and explains the higher discrepancy observed for these quantities at ignition.

At stagnation all the quantities are correctly captured by \texttt{FLAIM}, except the pusher density which is a factor of $\sim 3$ lower. However, the pusher density $\rho_\mathrm{p}\approx 2000~\unit{g/cm^3}$ mentioned in the reference paper  \cite{Molvig_PhysPlasmas_2018} is not entirely consistent to what reported in the corresponding plot at stagnation, in which a narrow peak at $\rho_\mathrm{p}\approx 2500~\unit{g/cm^3}$  is visible, suggesting that a lower \textit{average} density would be probably more accurate ($\rho_\mathrm{p}\approx 1000/1500~\unit{g/cm^3}$) and better in line with our results.

At peak burn the ion and electron temperatures separate and show very good agreement with Revolver, as well as the energy distribution between fuel and pusher. We measure a relative error of $\sim 10-17\%$ for all variables at peak burn, which directly impact the prediction on the final burn metrics. Regarding this, \texttt{FLAIM} predicts a final neutron energy yield of $E_{\mathrm{n}}=3.87~\unit{MJ}$ and a fuel burn fraction  of $\Phi_\mathrm{f}= 0.57$. Molvig et al. report a burn-up fraction of $\Phi_\mathrm{f}^{\mathrm{ref}}\approx 0.5$, in reasonable agreement. Given that the maximum possible neutron energy yield from Revolver (assuming complete burn) is $E_\mathrm{n}^{\mathrm{max}}=6.75$ MJ, the value of $\Phi_\mathrm{f}^{\mathrm{ref}}$ provides an estimated reference neutron energy yield $E_\mathrm{n}^{\mathrm{ref}}=3.4$ MJ, which is comparable with our result ($\epsilon \approx 0.13$). This is consistent with the minor  overestimation of $E_\mathrm{n}$  reported in Section \ref{sec:multi_physics_simulations} (Figure \ref{pic:neutron_yield_heat_map}) and can be attributed to the simplified treatment of the heat losses in \texttt{FLAIM}, in particular thermal conduction, which dominates the heat losses in the burning regime  (Section \ref{subsec:heat_losses}).

In conclusion, we find that \texttt{FLAIM} does indeed give suitably accurate results for such systems to be valuable as a predictive design tool, with the vast bulk of the physics captured by numerically inexpensive and simple models. This gives us confidence in our ability to use \texttt{FLAIM} as an accurate and robust surrogate for integrated simulations of volume ignition systems that we can leverage in large-scale optimisation and parameter space exploration.

\section{Conclusions}
\label{conclusions}
\noindent In this work we have presented \texttt{FLAIM}, a reduced model for the compression and thermonuclear burn of a spherical DT fuel capsules with high-Z pushers, in the context of the volume ignition scheme. \texttt{FLAIM}'s modularity and extensibility lend it significant utility in the performance of rapid investigations and it is used extensively at FLF for global optimisation, sensitivity analyses and large parameter scans. \texttt{FLAIM} is computationally very fast (\ref{app:spatial_convergence}), with typical \replaced[id=P2]{simulations with all operators activated}{full-physics simulations} running in $\sim 2-8$ s (with potential margin for improvement). This has allowed us to execute large optimisation campaigns comprised of tens of thousands of simulations, obtaining relevant results in just a few hours.

The objective of this work was to present \texttt{FLAIM} from the point of view of the physical and numerical modelling. \texttt{FLAIM} relaxes many of the assumptions commonly adopted in the reduced modelling of the volume ignition in ICF capsules, employing (i) hydrodynamic models independent of the type of EoS used as a closure; (ii) tabulated EoS and transport properties, accounting for the non-ideality typical of extreme physical conditions; (iii) a model for the wall able to track the radiation diffusion wave in the pusher based on the heat flux from the fuel; and (iv) a detailed thermonuclear burn operator which accounts for reactivity reduction, alpha-particle escape and pusher heating, as well as fuel depletion. Finally, we have implemented an additional hydrodynamic operator based on a 1D Lagrangian description of the fuel and pusher regions. This has been done to improve on the commonly used  assumption of infinitesimally thin or constant-thickness pushers. The impact of this modification on the computational cost of a \texttt{FLAIM} simulation has been found to be very limited (Figure \ref{pic:spatial_convergence}d), given the inherent capability of Lagrangian descriptions of high convergence ratio implosions to provide very reliable results with low resolution. Moreover, the benefits in terms of physical accuracy outweigh the (minimal) cost. We compared pure hydrodynamic simulations using \texttt{FLAIM} to our in-house hydro code \texttt{B2} over a large parameter space, demonstrating that for relatively thick pushers an accurate description of the hydrodynamics is essential. A constant thickness pusher (or an infinitesimally thin one) greatly overestimates the compressive  work done on the capsule, consuming all of its kinetic energy to increase the internal energy of the fuel. Whilst this assumption is reasonable for very thin pushers, it rapidly worsens the predictive capabilities of the model even for moderately thick ones. Conversely, a 1D description of the hydrodynamics (even using a few cells) is able to correctly predict the compression work and the energy distribution between the fuel and the shell with a negligible increase in simulation run-time.

Another major advantage of considering a 1D description \textit{only} for the hydrodynamics, is that all the other operators, that could be very complex to implement in a spatially-dependent framework, can be carried out as simple 0D models. An obvious example is the burn operator, for which a full 1D description would also require equations to govern the transport of species ($\alpha$ particles in particular) across the whole domain. In \texttt{FLAIM} this is not necessary and simple models available in literature can be used to estimate the fraction of particles that leave the fuel region based on the capsule size. Contrary to what we observed for the hydrodynamics, the impact of this approach for the burn physics is actually very limited and entirely justifiable. To support this, we ran a \replaced[id=P2]{simulation scan (including hydrodynamics, burn and losses)}{full-physics (i.e., including hydro, burn and losses) simulation scan} comparing the results obtained from \texttt{FLAIM} and \texttt{B2}, showing \replaced[id=P2]{good}{excellent} agreement in terms of final neutron yield and burn fraction in the whole parameter space. We found that the residual discrepancies are attributable to the way the heat losses are treated in the model. Specifically, we found the approximate model for thermal conduction to govern the \texttt{FLAIM}-\texttt{B2} discrepancy in the main ignition region, where we know the conductive fluxes dominate over radiation. Conversely, our simplified radiation model mainly impacts the boundaries of the ignition region, but not the accuracy in terms of burn performance metrics (i.e., neutron yield and burn fraction). 

Finally, we used \texttt{FLAIM} to model the Revolver design, showing good agreement with literature results for some selected key variables evaluated at ignition, stagnation and peak burn. For the latter, we report errors of $\sim 10-17\%$ for all the quantities investigated, increasing our assurance in the reliability of \texttt{FLAIM} as an accurate surrogate for predicting target performance in the peak burn phase.

Future work will focus on closing the existing gap between \texttt{FLAIM} and 1D high-fidelity simulations, mainly focusing on the heat loss models. Further optimisations include the implementation of alternative solutions for the integration of the ODE system, as well as improvements regarding the computational cost and residual code overhead.

\newpage
\appendix

\section{Spatial convergence}
\label{app:spatial_convergence}
\noindent The Revolver design \cite{Molvig_PhysRevLett_2016} presented in Section \ref{sec:revolver} is run with \texttt{FLAIM} at different spatial resolutions to evaluate the convergence behaviour. The error is evaluated based on the values of the final neutron yield $Y_\mathrm{n}$ and the maximum ion temperature $T_\mathrm{i}$. The value at zero grid spacing is estimated by the  Richardson extrapolation \cite{roache1993completed} of the quantities of interest. Using a \textit{constant} refinement ratio $r$ ($r=2$ in our case), the observed order of convergence $p$ is

\begin{equation}
	p_y = \frac{\ln\left(\frac{y_3-y_2}{y_2-y_1}\right)}{\ln\left(r\right)} \quad \mathrm{for}~y \in \{Y_\mathrm{n}, T_{\mathrm{i}}\}\,,
\end{equation}

\noindent where $y_1, y_2, y_3$ are values of the metric $y$ at three \textit{decreasing} grid resolutions. If $p>0$ (i.e., the solution is converging), the Richardson extrapolation value $y_{0}$ can be computed as

\begin{equation}
	y_{0} \approxeq y_{\mathrm{Rich}} = y_1 + \frac{y_1-y_2}{r^p - 1} \,,
\end{equation}

\noindent and the estimated relative error for a number of simulations  $N_{\mathrm{sims}}\geq3$ is computed as

\begin{equation}
	\epsilon_i = \frac{|y_i-y_0|}{y_0} \quad \mathrm{for}~i=0,\dots, N_{\mathrm{sims}}\,.
\end{equation}

\begin{figure}
	\centering
	\subfloat[]
	{\includegraphics[width=.49\textwidth,height=0.19\textheight]{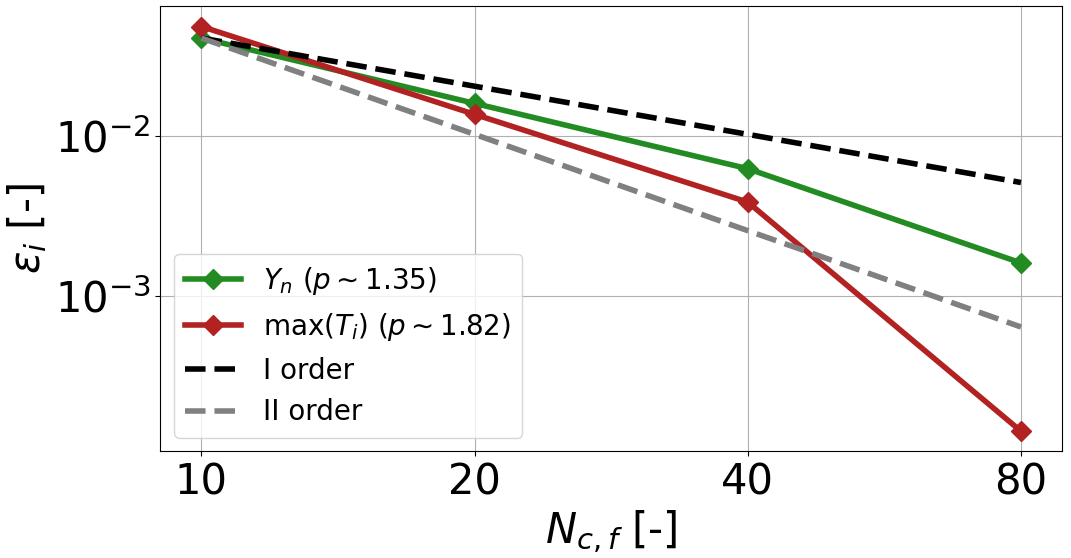}} ~
	\subfloat[]
	{\includegraphics[width=.49\textwidth,height=0.19\textheight]{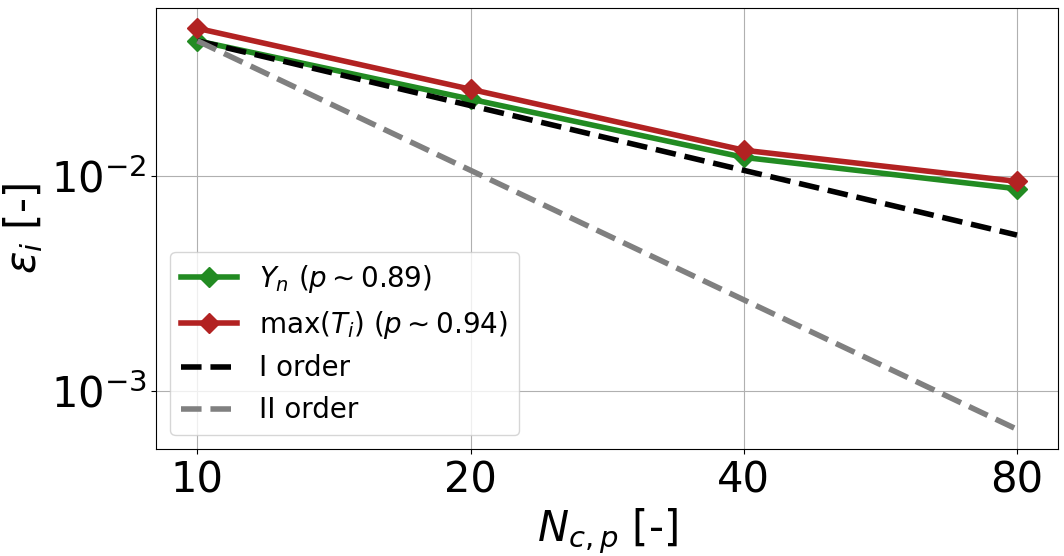}} \\
	\subfloat[]
	{\includegraphics[width=.49\textwidth,height=0.19\textheight]{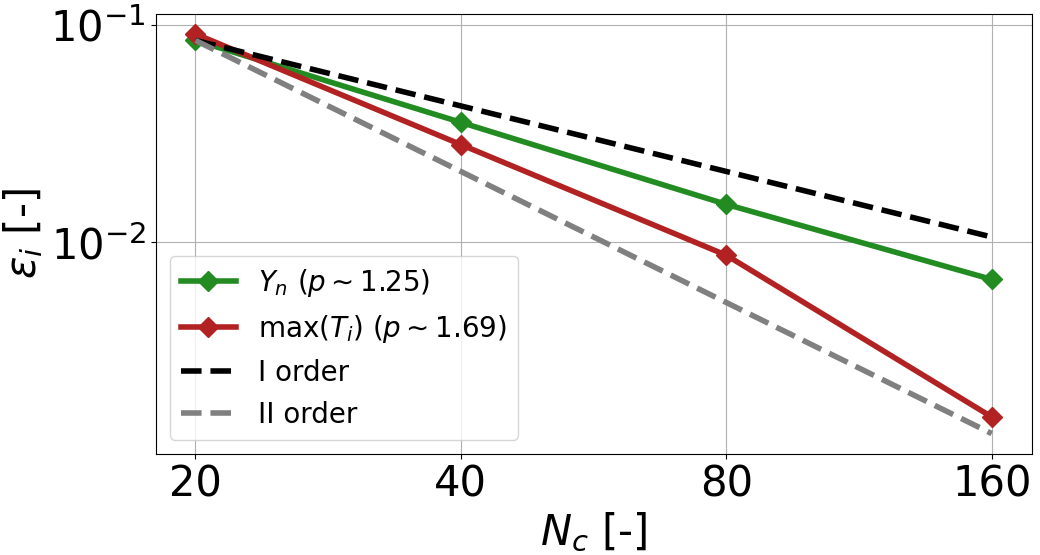}} ~
	\subfloat[]
	{\includegraphics[width=.49\textwidth,height=0.19\textheight]{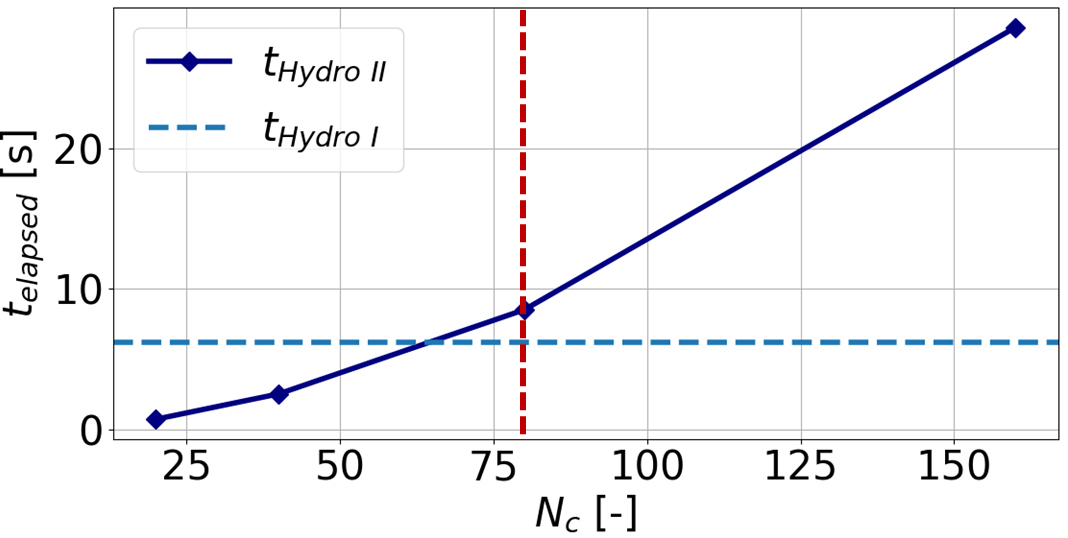}} 
	\caption{Grid convergence analysis on the number of cells in the fuel (a) (with $N_{\mathrm{c, p}}=80$), on the number of cells in the pusher (b) (with $N_{\mathrm{c, f}}=80$), on the total number of cells $N_\mathrm{c}$ (c). The observed orders of convergence are  reported in the legend for $Y_\mathrm{n}$ and $\mathrm{max}\left(T_\mathrm{i}\right)$. Total elapsed time is shown in (d) for the simulations shown in (c) run with models Hydro I, II. The vertical dashed red line in (c) indicates the resolution $N_\mathrm{c}=80$ typically used for \texttt{FLAIM} simulations.}
	\label{pic:spatial_convergence}
\end{figure}

\noindent Figure \ref{pic:spatial_convergence} reports the plot of $\epsilon_i$ vs. the number of cells\footnote{The size of the grid cells is not relevant for this analysis, since it is not constant for moving grids.} in the fuel (a), in the pusher (b) and total (c).

All simulation metrics show good convergence properties, with orders $0.9 \lesssim p \lesssim 2$. The observed order of convergence is affected by the time stepping scheme, currently first order, as well as by the presence of hydrodynamic discontinuities, i.e. shocks. Interestingly, the convergence order is different when refining the fuel and the pusher region independently, whereas an intermediate order is observed when both are changed at the same time. 

The elapsed time (Figure \ref{pic:spatial_convergence}d) remains below $30~\unit{s}$ for the most resolved case. Using  $N_{\mathrm{c}}=80$, we can obtain well-resolved (and converged) simulations ($\epsilon_i \approx 0.01$) at a cost ($t_{\mathrm{elapsed}} \approx 8~\unit{s}$) comparable to what provided by the non-discretised model Hydro I ($t_{\mathrm{elapsed}}~\approx 6~\unit{s}$). To give a reference, an equivalent converged 1D \texttt{B2} simulation of the Revolver design  showed $t_{\mathrm{elapsed}}~\approx 3~\unit{h}$. For the scan in Table \ref{table:b2_scan}, all 125 simulations were launched in parallel for the two codes, with an elapsed time of $\sim 5~\unit{h}$ for \texttt{B2} and $\sim 10~\unit{s}$ for \texttt{FLAIM} (both governed by the slowest simulation),  highlighting the advantages of using \texttt{FLAIM} for quickly run large parameter scans and sensitivity analyses that often require thousands of simulations.
\clearpage

\section{Operator ordering}
\label{app:operator_splitting}
\noindent When adopting an operator splitting approach for the solution of an ODE system, one carries numerical errors due to the decoupling on the physical operators. Moreover, it is inevitable to obtain different solutions for different choices of the operator ordering. However, one must ensure that this discrepancy vanishes for increasing resolution of the time discretisation, effectively leading to a code that produces consistent results independent of the specific order of the physical operators involved. To this purpose, we ran a simple test:

\begin{figure}
	\centering
	{\includegraphics[width=.54\textwidth,height=0.23\textheight]{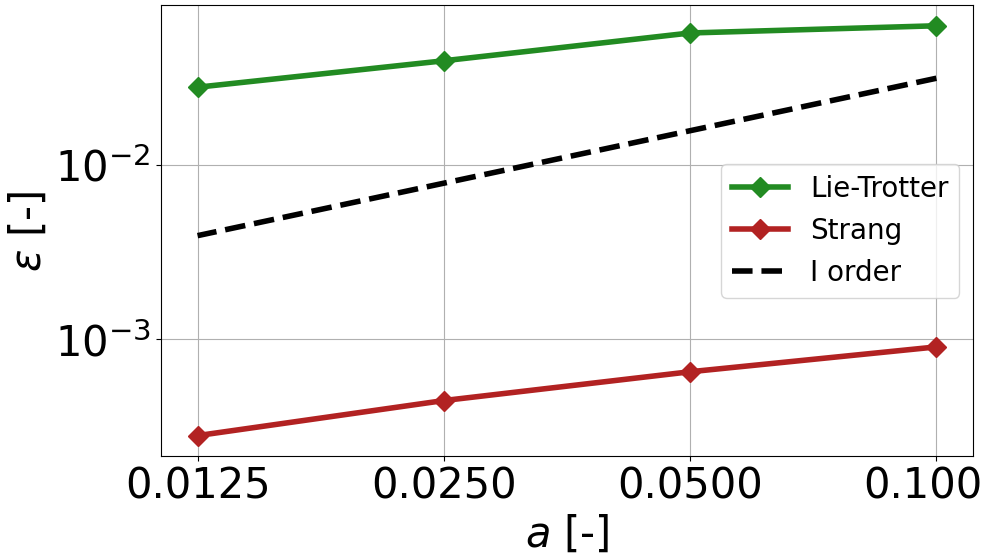}}
	\caption{Convergence analysis for the average distance $\epsilon$ between the $T_\mathrm{i}(t)$ profiles generated by different physical  operator ordering (Equation \ref{operator_splitting_eq}).}
	\label{pic:operator_splitting}
\end{figure}

\begin{itemize}
	\item The Revolver design (Section \ref{sec:revolver}) is run   $N_{\mathrm{sims}}$ times,  with different randomly generated operator ordering ($N_{\mathrm{sims}}=4$). The physical operators are 6, i.e. hydrodynamics, thermal-equilibration, conduction, radiation, wall, burn (Section \ref{section:mathematical_model});
	\item For these $N_{\mathrm{sims}}$ simulations, the error $\epsilon$ (interpreted as the average distance between them) is calculated by means of the average of the $L_2$ norms of the ion temperature profiles $T_\mathrm{i}(t)$ (taking one of them as the reference) using 
		 \begin{equation}
		 		\begin{split}
				\epsilon &= \\
				&=\frac{1}{N_{\mathrm{sims}}}\sum_{j}^{N_{\mathrm{sims}}}L_2\left(T_{\mathrm{i}, j}\right) \\
				&=\frac{1}{N_{\mathrm{sims}}}\sum_{j}^{N_{\mathrm{sims}}} \sqrt{\sum_{k}^{N_{\mathrm{steps}, j}}\left(\frac{T_{\mathrm{ref}}\left(t_k\right) - T_{\mathrm{i}, j}\left(t_k\right)}{T_{\mathrm{ref}}\left(t_k\right)}\right)^2}  \quad\quad \,;
				\end{split}
				\label{operator_splitting_eq}
		 \end{equation}
	 \item The procedure is applied at decreasing Courant numbers $a$ (Equation \ref{cfl_condition}) to verify the convergence properties of $\epsilon$.
\end{itemize}

\noindent The analysis is done for both the Lie-Trotter and the Strang splitting presented in Section \ref{section:numerical_setup} and Figure \ref{pic:operator_splitting} summarises the results. The average value $\epsilon$ is in general very low   ($10^{-1}\lesssim \epsilon \lesssim 10^{-4}$) and, as expected, shows the lowest values for the Strang splitting. The different profiles $T_\mathrm{i}(t)$ converge with time resolution, allowing us to conclude that \texttt{FLAIM} results are effectively independent of the operator ordering.
\clearpage

\section{\added[id=P9]{The B2 code}}
\label{app:b2_code}
\noindent In this paper the \texttt{B2} code is extensively used as a benchmark code for assessing the reliability of \texttt{FLAIM} as a reduced volume ignition model. \texttt{B2} is one of FLF's main production codes for the design of our targets and it is continuously and extensively verified against a suite of benchmark test-cases and validated with in-house and external experimental results. We present here a brief description of the main physical models implemented in \texttt{B2}, as well as some selected verification tests relevant to support this work. \\
\noindent \texttt{B2} is an Eulerian 3D parallel multi-material RMHD (Resistive-Magneto-Hydrodynamic) code adopting structured orthogonal grids. The hydrodynamics is based on a  Lagrangian remap approach with artificial viscosity, coupled with a geometric Volume Of Fluid (VOF) methodology for the interface capturing \cite{hirt1981volume} (using a Simple Line Interface Calculation (SLIC) method for the local interface reconstruction).  Several radiation transport models are available (multi-group radiation transport, with $\mathbb{P}_{1/3}$ or radiation diffusion closure and an exact 6D Monte Carlo transport), whilst thermal conduction is applied to both ion and electron species, accounting for flux-limiting. The thermonuclear burn model includes fuel depletion and single-group Eulerian $\alpha$ particle diffusion to describe  the fusion self-heating \cite{atzeni1981diffusive}. EoS and transport properties are tabulated for temperatures up to $T_\mathrm{i}=100~\unit{keV}$ and, since they are shared between $\texttt{B2}$ and $\texttt{FLAIM}$, the details about  their calculation have been already reported in Section \ref{section:mathematical_model}.

\subsection{Hydrodynamics}
\noindent Lagrangian remap \cite{arber2001staggered} is adopted to model the hydrodynamics, solving the generalised two-temperature Euler equations for mass, momentum and total energy conservation. The hydrodynamic scheme is based on a first-order Godunov formulation (which is the one used in this work), but higher-order reconstruction schemes (up to third-order) are also available. We report two classic verification tests for hydrodynamics: (i) Sod shocks \cite{sod1978survey} and (ii) Noh implosions \cite{noh1987errors}.

\begin{figure}
	\centering
	{\includegraphics[width=1\textwidth,height=0.9\textheight]{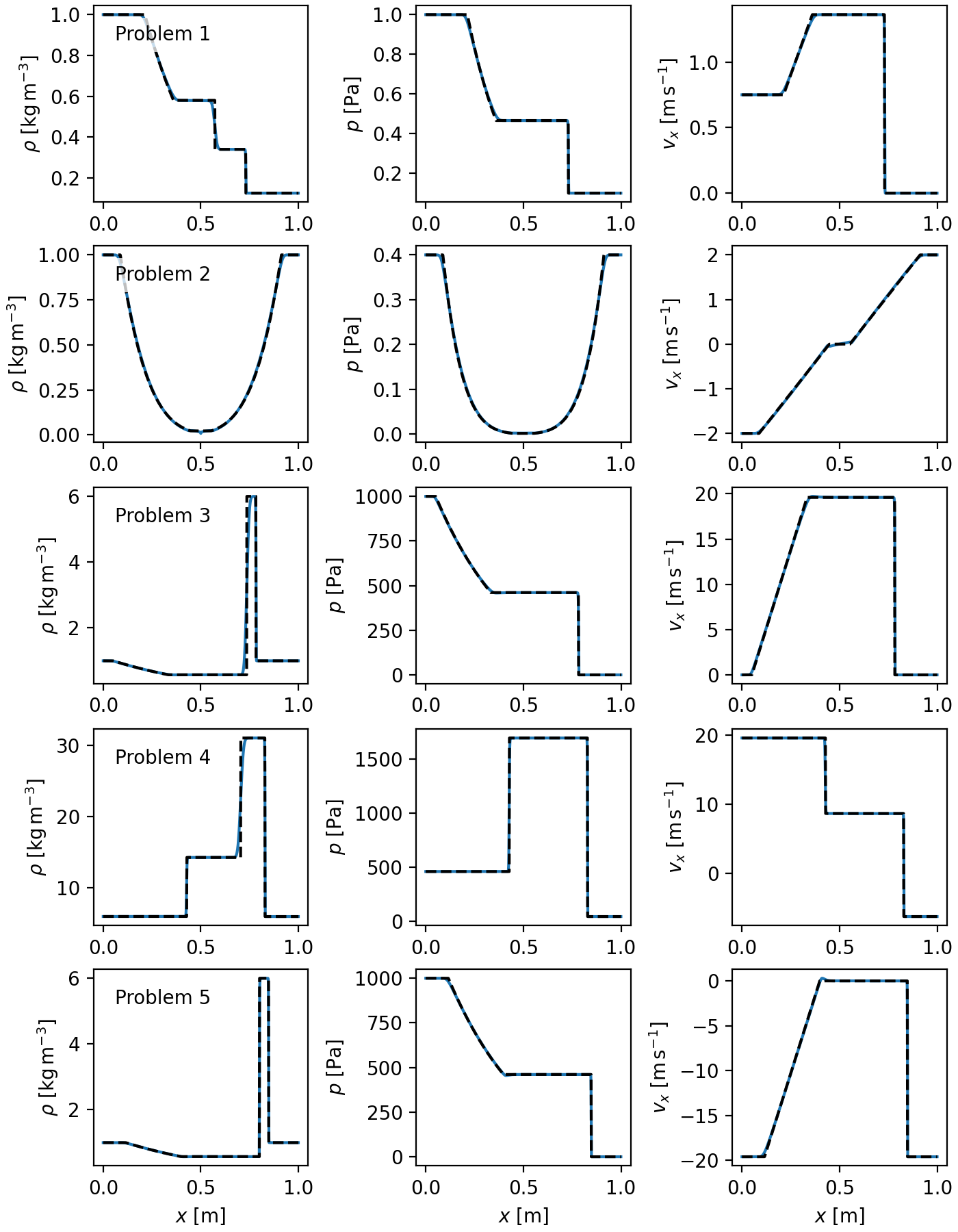}}
	\caption{Sod shocks: comparison between \texttt{B2} (blue solid line) and the analytic solution (black dashed line) in terms of spatial profiles of mass density $\rho$, pressure $p$ and fluid velocity $v_\mathrm{x}$ for the 5 problems reported in Toro et al. \cite{toro2013riemann}. For all cases, $n_\mathrm{c}=2048$ and $1^\mathrm{st}$ order Godunov scheme is used.}
	\label{pic:sod_shocks}
\end{figure}

\subsubsection{Sod shocks}
\begin{figure}
	\centering
	{\includegraphics[width=1\textwidth,height=0.3\textheight]{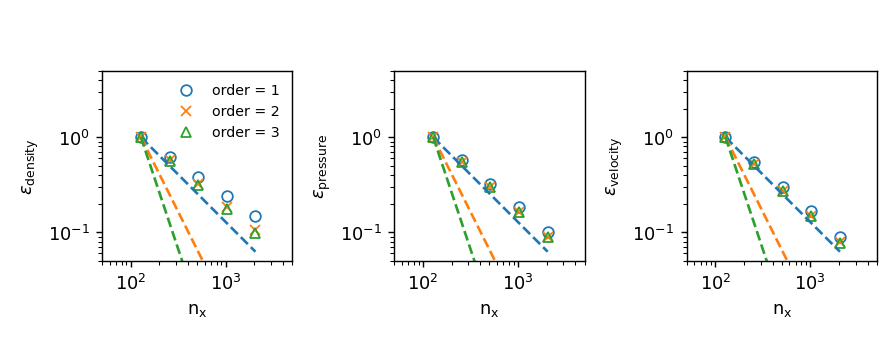}}
	\caption{Sod shocks: grid convergence properties of \texttt{B2} for Sod problem 1 (Figure \ref{pic:sod_shocks}, top row), using three different hydrodynamic schemes with $1^{\mathrm{st}}$,  $2^{\mathrm{nd}}$ and $3^{\mathrm{rd}}$ theoretical order of accuracy.}
	\label{pic:sod_shocks_convergence}
\end{figure}

\noindent This is a standard test for hydrodynamic solvers, consisting of a 1D Riemann problem that, when evolved in time, produces five distinct regions with a shock, a rarefaction and a contact discontinuity. The five test cases presented in this Appendix are  extensively described in Toro et al. \cite{toro2013riemann} (using an adiabatic index $\gamma=5/3$), and the results are reported in Figure \ref{pic:sod_shocks}. Convergence properties of the \texttt{B2} code for the Sod 1 problem are shown in Figure \ref{pic:sod_shocks_convergence} using three hydrodynamic schemes at different orders of convergence (using a standard normalised $L_1$ norm to measure the error $\epsilon$). The discontinuous nature of the solution and the first-order time stepping scheme limit the observed convergence to be $\sim1^{\mathrm{st}}$ order, as expected.

\subsubsection{Noh implosions}
\begin{figure}
	\centering
	{\includegraphics[width=0.9\textwidth,height=0.35\textheight]{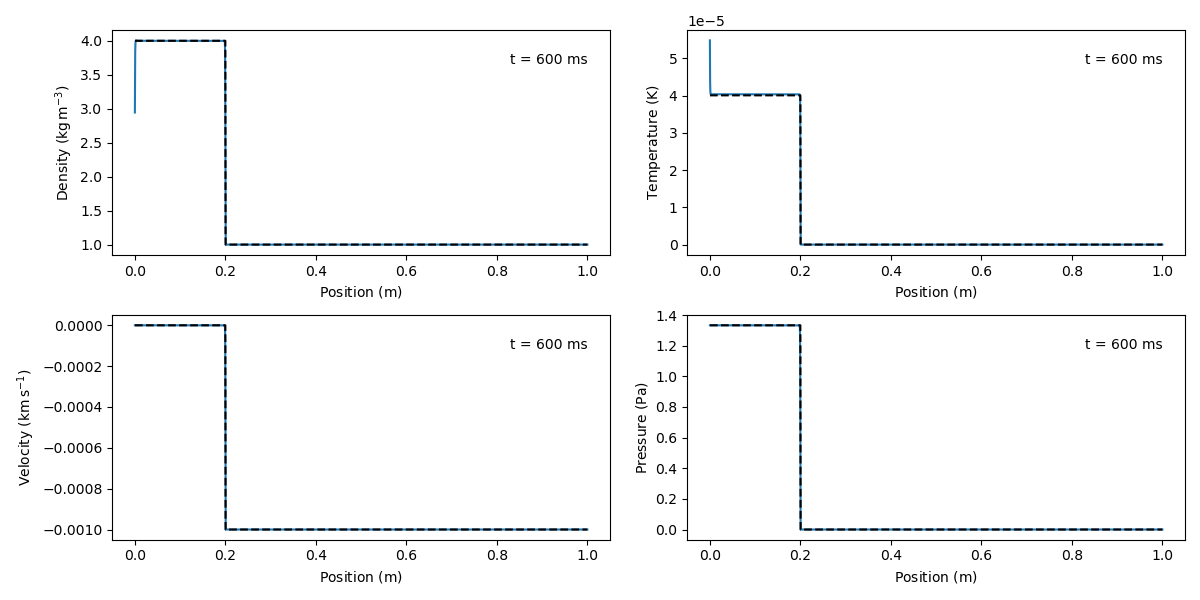}}
	\caption{Noh implosion (planar) \cite{noh1987errors}: mass density, temperature, velocity and pressure spatial profiles, comparison between \texttt{B2} (blue solid line, $n_\mathrm{c}=2048$) and the analytic solution (black dashed line).}
	\label{pic:noh_implosions_planar}
\end{figure}

\begin{figure}
	\centering
	{\includegraphics[width=0.9\textwidth,height=0.35\textheight]{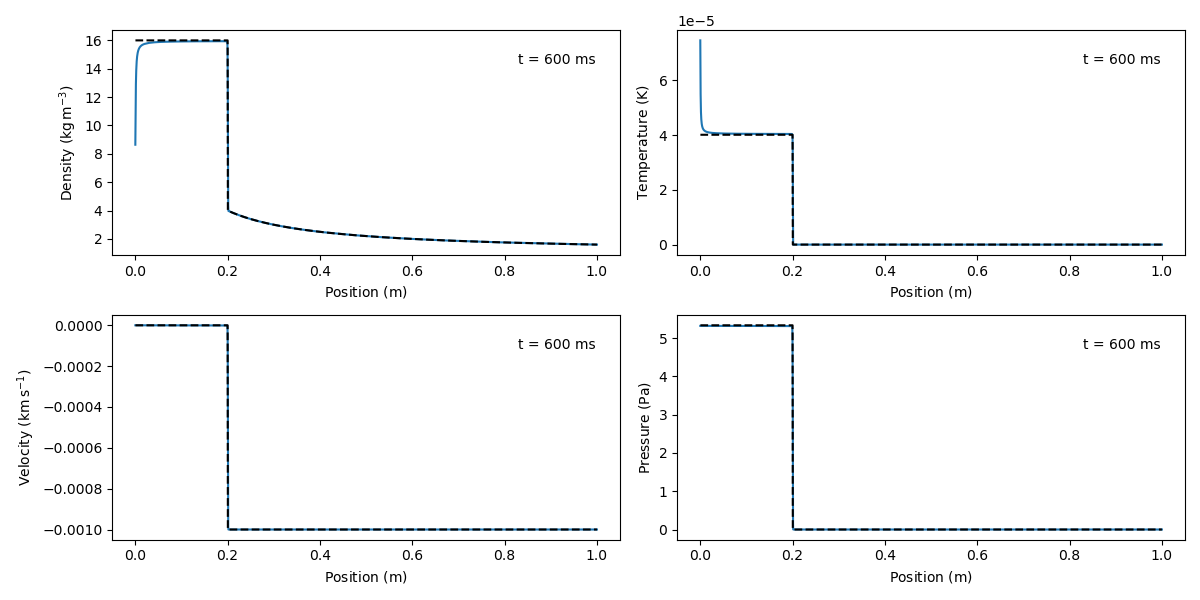}}
	\caption{Noh implosion (axial) \cite{noh1987errors}: mass density, temperature, velocity and pressure spatial profiles, comparison between \texttt{B2} (blue solid line, $n_\mathrm{c}=2048$) and the analytic solution (black dashed line).}
	\label{pic:noh_implosions_axial}
\end{figure}

\begin{figure}
	\centering
	{\includegraphics[width=0.9\textwidth,height=0.35\textheight]{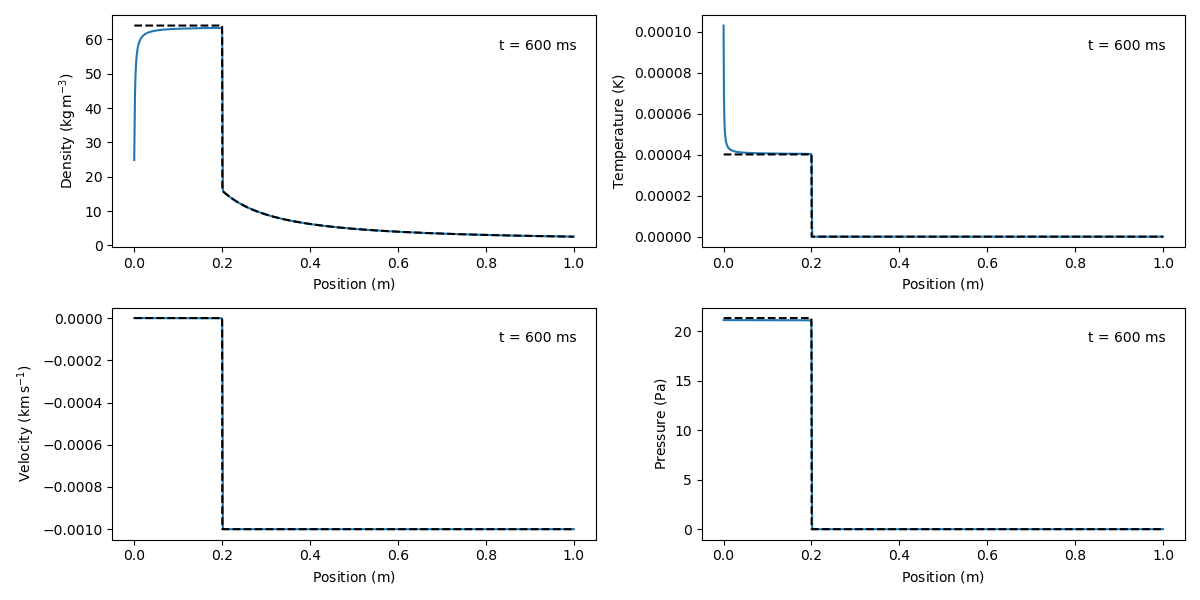}}
	\caption{Noh implosion (spherical) \cite{noh1987errors}: mass density, temperature, velocity and pressure spatial profiles, comparison between \texttt{B2} (blue solid line, $n_\mathrm{c}=2048$) and the analytic solution (black dashed line).}
	\label{pic:noh_implosions_spherical}
\end{figure}

\begin{figure}
	\centering
	\subfloat[]
	{\includegraphics[width=.45\textwidth,height=0.25\textheight]{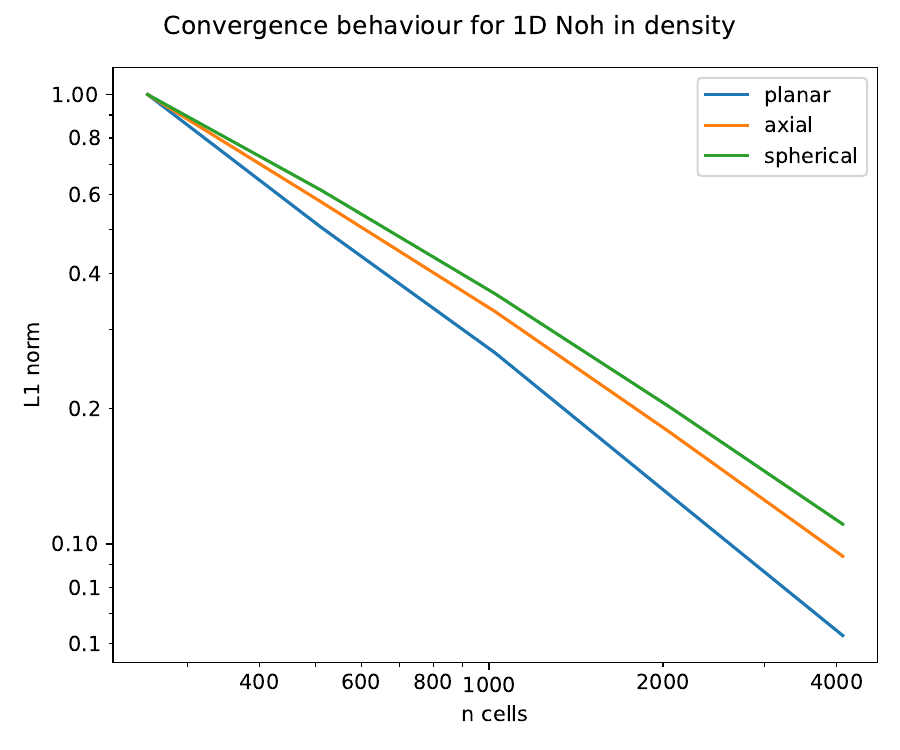}} ~
	\subfloat[]
	{\includegraphics[width=.45\textwidth,height=0.25\textheight]{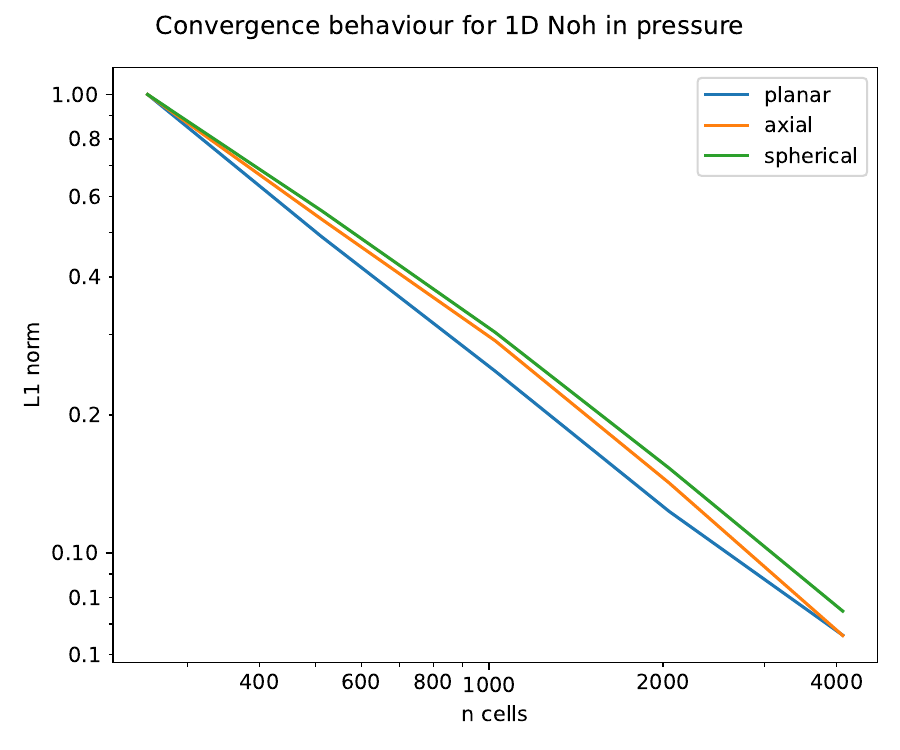}}
	\caption{Noh implosions: convergence properties of \texttt{B2} for planar, axial and spherical geometry.}
	\label{pic:noh_implosion_convergence}
\end{figure}

\noindent The Noh implosion test case \cite{noh1987errors} exercises hydrodynamics, convergent geometry and implosion physics in planar, axial or spherical geometry. In a domain $0 \leq x \leq 1$, a cold ideal gas ($\gamma=5/3$) at initially zero pressure and temperature falls inwards at constant velocity ($v=-1~\unit{m/s}$) and stagnates on the axis, producing an expanding shock front propagating into the collapsing gas. It is known to be a very challenging test for ICF codes, with most codes reporting errors in the form of wall heating, lack of symmetry, incorrect shock speeds or even failure to run \cite{ramani2023fast}. Results are shown in Figures \ref{pic:noh_implosions_planar}, \ref{pic:noh_implosions_axial}, \ref{pic:noh_implosions_spherical}  for a planar, axial and spherical geometry respectively. The plots show good agreement with the analytic results, except for the region near the axis, where classic wall overheating can be noticed \cite{rider2000revisiting}. As for the Sod shock test cases, the convergence analysis is carried out in terms of standard normalised $L_1$ norm and it is reported in Figure \ref{pic:noh_implosion_convergence}, showing $\sim 1^{\mathrm{st}}$ order convergence.

\subsection{Thermal conduction}
\noindent A standard flux-based finite-volume formulation for ion and electron thermal conduction is implemented in \texttt{B2}, augmented with flux-limiters to account for the upper limit on the heat flux given by the particle thermal speed. The benchmark test cases we present here are (i) a simple thermal relaxation test and (ii) the Zel'dovich and Raizer thermal wave verification test \cite{Zeldovich_book}.

\subsubsection{Thermal relaxation}
\noindent This test models the relaxation of a small Gaussian temperature perturbation. For a single temperature system with constant thermal conductivity, a closed-form solution can be derived. A 1D domain $\left[-10 ~ \unit{mm} \leq x \leq 10 ~ \unit{mm}\right]$, with a constant background density, has a temperature profile initialised as

\begin{equation}
	T\left(x\right)=T_0 + T_1 e^{-\left(\frac{x}{w}\right)^2}
	\label{eq:temperature_perturbation_test} \,,
\end{equation}

\noindent with $T_0=100~\unit{eV}$, $T_1=10~\unit{eV}$ and $w=1~\unit{mm}$. The thermal conductivity is set to $\kappa_0=57000~\unit{W/m/K}$ (this approximates the Spitzer electron thermal conductivity of a fully ionised helium plasma at $T = T_0$), whilst the volumetric heat capacity is fixed to $c_{\mathrm{v}}=520~ \unit{J/m^3/K}$. The analytic solution for the 1D transient heat diffusion equation is

\begin{equation}
	T\left(x, t\right)=T_0 + T_1\left(\frac{w}{w_0}\right)e^{-\left(\frac{x}{w_0}\right)^2} \,,
\end{equation}

\noindent where $w_0=\sqrt{w^2 + 4\kappa_0 t / c_{\mathrm{v}}}$. The comparison between the $\texttt{B2}$ simulation and the analytic solution is reported in Figure \ref{pic:thermal_conduction_tests}a.

\subsubsection{Zel'dovich-Raizer thermal wave}
\begin{figure}
	\centering
	\subfloat[]
	{\includegraphics[width=.42\textwidth,height=0.2\textheight]{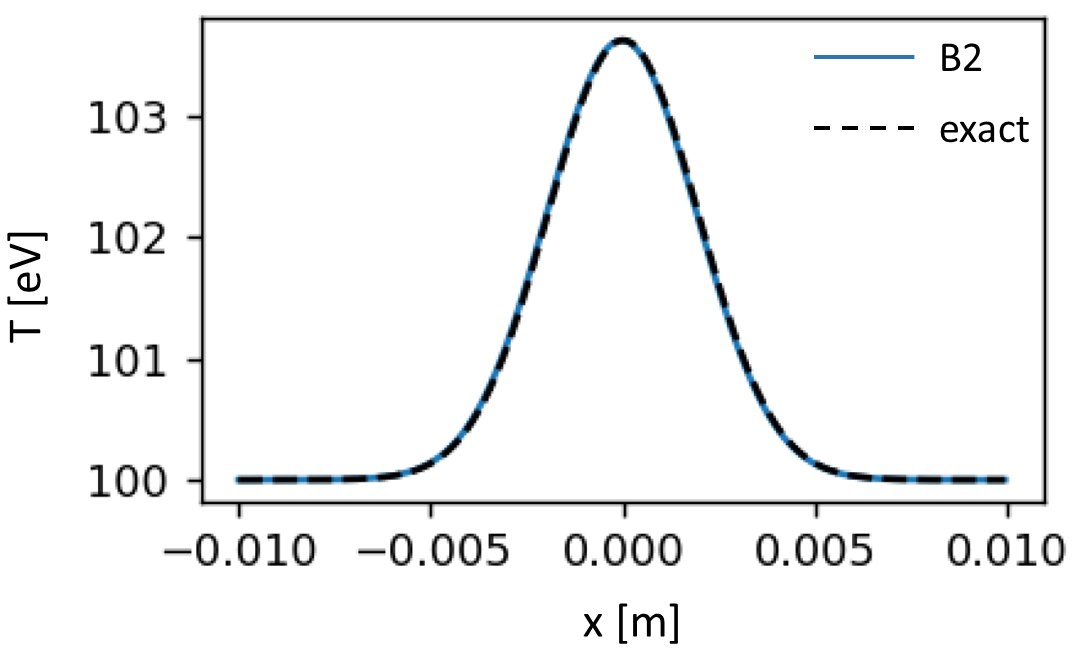}} ~
	\subfloat[]
	{\includegraphics[width=.45\textwidth,height=0.2\textheight]{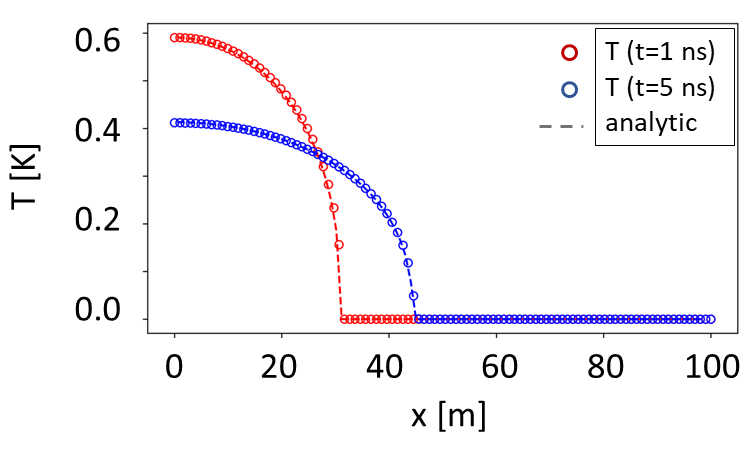}}
	\caption{Thermal conduction verification tests: thermal relaxation (a) at $t=15~\unit{ns}$ and Zel'dovich-Raizer thermal wave (b), both showing the spatial profiles of temperature obtained with \texttt{B2} vs. the corresponding analytic solution.}
	\label{pic:thermal_conduction_tests}
\end{figure}

\noindent This test is based on a self-similar solution to thermal wave propagation in a cold medium with a non-linear thermal conductivity, as described in Chapter 10 of Zel'dovich and Raizer's textbook \cite{Zeldovich_book}. The domain spans $0 \leq x \leq 100~\unit{\mu m}$, the EoS is deuterium ideal gas ($\gamma=5/3$) and the thermal conductivity depends on temperature following $\kappa=\zeta c_\mathrm{v} T^n$ (with $c_\mathrm{v}$ being the volumetric heat capacity, $\zeta=1~\unit{m^2/s/K^n}$ and $n=5/2$). The analytic solution is given by Eqs. (10.32)-(10.33) in \cite{Zeldovich_book}. The wave dynamics is reported in Figure \ref{pic:thermal_conduction_tests}b in terms of spatial temperature profiles at different time instants.

\subsection{Radiation diffusion}
\noindent As reported in the introduction of this section, several models are available in \texttt{B2} for modelling radiation transport. In this work, the standard radiation diffusion approach \cite{Pomraning_book} has been used for the comparisons in Section \ref{sec:multi_physics_simulations} in order to be consistent with the wall model \cite{Dodd_PhysPlasmas_2020, Hammer_PhysPlasmas_2003} implemented in \texttt{FLAIM} (which provides a semi-analytic solution for a 1D radiation diffusion equation in the shell). The McClarren solution is adopted for verification purposes \cite{mcclarren2021two}.

\begin{figure}
	\centering
	{\includegraphics[width=0.75\textwidth,height=0.6\textheight]{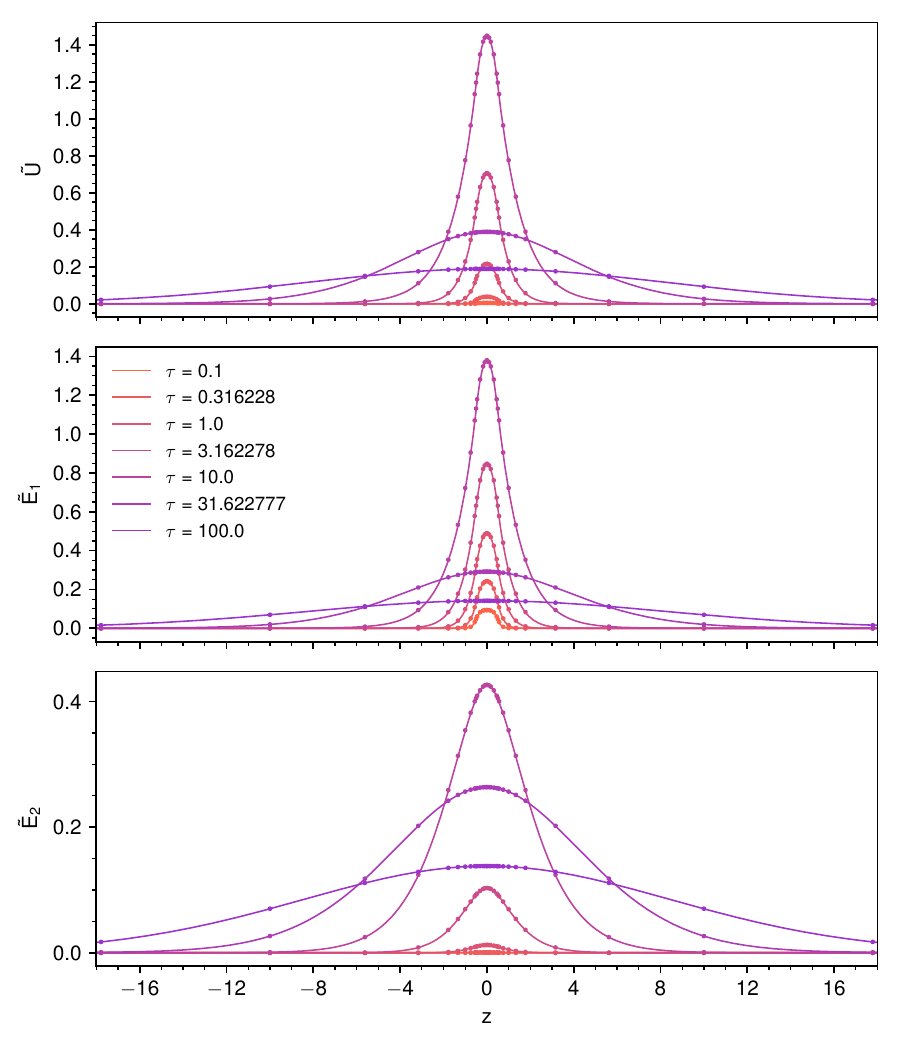}}
	\caption{McClarren test case \cite{mcclarren2021two}: $\tilde{E}_i=E_i/\left(aT_\mathrm{H}^4\right)$ is the normalised radiative energy densities for group $i$, whilst $\tilde{U}=U/\left(aT_\mathrm{H}^4\right)$ is the normalised internal energy density of the medium. The holraum temperature is $T_\mathrm{H}=1~\unit{keV}$. Analytic solution (dots) vs. \texttt{B2} simulation (solid line).}
	\label{pic:mcclarren_test}
\end{figure}

\subsubsection{McClarren test case}
\noindent McClarren  \cite{mcclarren2021two} provides an analytic solution for a two-group Marshak wave with spatially constant opacities $\chi_1, \chi_2$. Given a  symmetric 1D domain, a line-source of radiation is applied to a fixed region $-z_0/2 \leq z \leq z_0/2$ around the origin for a short time ($\tau<\tau_0$, with $\tau=c\chi_1 t$), after which it is switched off, letting the radiation diffuse towards the boundaries. The test verifies that (i) two radiation groups can be transported correctly under the diffusion approximation and that (ii) the radiative energy density of the groups is correctly coupled with the internal energy density of the medium via absorption and emission. The results are reported in Figure \ref{pic:mcclarren_test}.



%%  \bibliography{<bibliography>}


\bibliographystyle{elsarticle-num}
\bibliography{papers,books}

\end{document}